\documentclass{emulateapj}
\usepackage{graphics,graphicx,epsfig}
\usepackage{amsmath,amssymb,amsfonts} 
\usepackage{subfigure, multirow}
\usepackage{epstopdf}
\usepackage{natbib}
\newcommand{\ergs}{erg s$^{-1}$}

\newcommand{\msun}{M$_{\odot}$~}
\newcommand{\msunyr}{M$_{\odot}$ yr$^{-1}$~}
\newcommand{\msunyrs}{M$_{\odot}$ yr$^{-1}$}

\newcommand{\eg}{\textit{e.g.,}~}
\newcommand{\ie}{\textit{i.e.,}~}

\newcommand{\HST}{\textit{HST} }

\newcommand{\Chandra}{\textit{Chandra} }
\defcitealias{L08}{L08}
\defcitealias{L05}{L05}
\defcitealias{X11}{X11}
\defcitealias{F12}{F12}

\begin{document}

\title{The X-ray Star Formation Story as Told by Lyman Break Galaxies in the 4~Ms CDF-S }
\author{ 
Antara R. Basu-Zych\altaffilmark{1},
Bret D. Lehmer\altaffilmark{1, 2, 3},
Ann E. Hornschemeier\altaffilmark{1}, 
Rychard J. Bouwens\altaffilmark{4, 5}, 
Tassos Fragos\altaffilmark{6}, 
Pascal A. Oesch\altaffilmark{5, 7}, 
Krzysztof Belczynski\altaffilmark{8,9},
W.N. Brandt\altaffilmark{10, 11},
Vassiliki Kalogera\altaffilmark{12}, 
Bin Luo\altaffilmark{10, 11}, 
Neal Miller\altaffilmark{13}, 
James R. Mullaney\altaffilmark{14}, 
Panayiotis Tzanavaris\altaffilmark{1}, 
Yongquan Xue\altaffilmark{15, 10, 11}, 
Andreas Zezas\altaffilmark{6, 16, 17}}
\altaffiltext{1}{NASA Goddard Space Flight Center, Code 662, Greenbelt, MD 20771; antara.r.basu-zych@nasa.gov}
\altaffiltext{2}{Department of Physics and Astronomy, The Johns Hopkins University, 3400 North Charles Street, Baltimore, MD 21218}
\altaffiltext{3}{Einstein Fellow}
\altaffiltext{4}{Leiden Observatory, Leiden University, NL-2300 RA Leiden, Netherlands}
\altaffiltext{5}{UCO/Lick Observatory, University of California, Santa Cruz, CA 95064}
\altaffiltext{6}{Harvard-Smithsonian Center for Astrophysics, 60 Garden Street, Cambridge, MA 02138, USA}
\altaffiltext{7}{Hubble Fellow}
\altaffiltext{8}{Astronomical Observatory, University of Warsaw, Al. Ujazdowskie 4, 00-478 Warsaw, Poland}
\altaffiltext{9}{Center for Gravitational Wave Astronomy, University of Texas at Brownsville, Brownsville, TX 78520}
\altaffiltext{10}{Department of Astronomy and Astrophysics, Pennsylvania State University, University Park, PA 16802, USA}
\altaffiltext{11}{Institute for Gravitation and the Cosmos, Pennsylvania State University, 525 Davey Lab, University Park, PA 16802, USA}
\altaffiltext{12}{Department of Physics and Astronomy, Northwestern University, 2145 Sheridan Road, Evanston, IL 60208, USA}
\altaffiltext{13}{Department of Astronomy, University of Maryland, College Park, MD 20742}
\altaffiltext{14}{Department of Physics, Durham University, South Road, Durham DH1 3LE, UK}
\altaffiltext{15}{Key Laboratory for Research in Galaxies and Cosmology, Department of Astronomy, University of Science and
Technology of China, Chinese Academy of Sciences, Hefei, Anhui 230026, China}
\altaffiltext{16}{University of Crete, Physics Department \& Institute of Theoretical \& Computational Physics, 71003 Heraklion, Crete, Greece}
\altaffiltext{17}{Foundation for Research and Technology-Hellas, 71110 Heraklion, Crete, Greece}

\begin{abstract}
We present results from deep X-ray stacking of $>$4000 high redshift galaxies from $z\approx$1 to 8 using the 4~Ms Chandra Deep Field South (CDF-S) data, the deepest X-ray survey of the extragalactic sky to date. The galaxy samples were selected using the Lyman break technique based primarily on recent \HST ACS and WFC3 observations.  Based on such high specific star formation rates (sSFRs): $\log$ SFR/$M_* >-8.7$, we expect that the observed properties of these LBGs are dominated by young stellar populations. 
The X-ray emission in LBGs, eliminating individually detected X-ray sources (potential AGN), is expected to be powered by X-ray binaries and hot gas. We find, for the first time, evidence of evolution in the X-ray/SFR relation. Based on X-ray stacking analyses for $z<4$ LBGs (covering $\sim 90$\% of the Universe's history), we find that the 2--10 keV X-ray luminosity evolves weakly with redshift ($z$) and SFR as: $\log$ L$_{\rm X}=0.93\log(1+z)+0.65\log {\rm SFR}+39.80$. By comparing our observations with sophisticated X-ray binary population synthesis models, we interpret that the redshift evolution of L$_{\rm X}$/SFR is driven by metallicity evolution in HMXBs, likely the dominant population in these high sSFR galaxies. We also compare these models with our observations of X-ray luminosity density (total 2--10~keV luminosity per $\rm Mpc^{3}$) and find excellent agreement. While there are no significant stacked detections at $z\gtrsim 5$, we use our upper limits from $5\lesssim z \lesssim 8$ LBGs to constrain the SMBH accretion history of the Universe around the epoch of reionization. 
\end{abstract}

\section{Introduction}\label{sec:intro}
Recently, Chandra completed the deepest X-ray survey to date: 4 Ms in the Chandra Deep Field-South \citep[CDF-S;][hereafter \citetalias{X11}]{X11}. The last leap forward in X-ray survey depth occurred approximately a decade ago as deep surveys moved from 1 Ms to 2 Ms coverage \citep[\eg][]{Alexander03}. Such deep X-ray surveys are crucial to studying the high-energy emission from distant star-forming galaxies, which overall are an extremely X-ray faint population compared to the more luminous active galactic nuclei (AGN) population that dominates the number counts at bright fluxes \citep[\eg see][]{Lehmer12}. Notably, since the profusion of papers several years ago on high redshift ($z>2$) X-ray studies of star-forming galaxies \citep[\ie][]{Reddy04, Laird05, Laird06, L05} there have been significant improvements in the identification of new high redshift galaxy samples, selected using the Lyman break technique, from $z=1.5-8$ \citep{Beckwith06, Bouwens07, Bouwens08, Bouwens10, Reddy09, Oesch10, Hathi10, Bunker10}. Thus, it is timely to study the evolution of X-ray properties in star-forming galaxies with these updated data sets. 

The Lyman break technique efficiently selects distant star-forming galaxies based on a strong spectral break at 912 \AA, caused by Lyman series absorption of neutral hydrogen in the intergalactic medium;
such galaxies are referred to as Lyman break galaxies \citep[LBGs;][]{Steidel92, Steidel93, Steidel95, Steidel00}. 
The color selection applied in searching for LBGs requires bright rest-frame UV continua and blue colors, therefore selecting galaxies with recent star formation and stellar populations dominated by young stars. 
LBGs have greatly impacted investigations of galaxy evolution. For example, significant effort has been devoted to studying the cosmic star formation history and particularly in measuring the peak (and decline at $z>3-4$) of the global star formation rate (SFR) density using LBGs \citep{Madau96, Bouwens06, Bouwens09, Bouwens11}. However, dust attenuation in UV-selected galaxies poses challenges to measuring the total SFR in these galaxies. Other samples of galaxies \citep[\eg submillimeter selected; ][]{Blain99} have been studied to provide additional constraints on the SFR density.

X-rays offer another way to explore the dust-unobscured cosmic star formation history. X-rays in ``normal galaxies'' (\ie not dominated by active galactic nuclei, AGN) mainly originate from accreting X-ray binaries (XRBs, including ultraluminous X-ray sources). The 2--10 keV emission in normal local galaxies scales with SFR \citep{Ranalli03, PR2007, Mineo12}. However a number of other sources can contribute to the X-ray emission in normal galaxies: supernovae and their remnants, and hot gas from starburst-driven winds and outflows \citep[see, \eg review by][]{Fabbiano89}. Low-luminosity AGN activity may also contribute to the X-ray luminosity in normal galaxies. 

Within the XRB population, high mass X-ray binaries (HMXBs) are short-lived, tracing recent star formation activity (on timescales $\sim10^{6-7}$ yrs), while low mass X-ray binaries (LMXBs) trace older stellar populations (for timescales $>10^{8-9}$ yrs) and the 2--10 keV X-ray luminosity scales with the stellar mass (M$_*$) of the galaxy. Assuming the following analytic parameterization for local (within a distance of 60 Mpc) normal galaxies:  
\begin{eqnarray}
L_{\rm X, 2-10 keV} & = & L_{\rm X}({\rm LMXB}) + L_{\rm X}({\rm HMXB}) \label{eqn:lx} \\
&=& \alpha{\rm M}_*  + \beta {\rm SFR} 
\end{eqnarray} 
\cite{Lehmer2010} measured constants of $\alpha=(9.05 \pm 0.37)\times 10^{28}$ \ergs~M$_\odot^{-1}$ and $\beta=(1.62\pm0.22) \times 10^{39}$ \ergs~ (\msunyr)$^{-1}$ \citep[see also][]{Colbert04}. 
As a relatively dust-insensitive probe of past and present SFR, the average X-ray properties of star-forming galaxies give an independent measurement of the cosmic star formation history. In addition, the wide range of evolutionary timescales, SFRs, and metallicities probed by deep surveys to high redshift offer constraints on binary evolution theories \citep[\eg][hereafter, referred to as \citetalias{F12}]{F12}. 

Several studies have shown the utility of stacking deep X-ray data at the optically determined locations of star-forming galaxies to measure the average X-ray properties of distant LBG populations that are individually undetected. Averaging the X-ray counts at the known positions of the targets enhances the signal-to-noise for source populations with fluxes below the detection limit. Through stacking, the relationship between X-rays and other galaxy properties (\eg SFR, stellar mass, dust attenuation) have been investigated for low redshift galaxies \citep[$z<1.4$; ][]{Ptak01, Hornschemeier02, Laird05, Lehmer07, L08, Watson09, Symeonidis11}, intermediate redshift LBGs \citep[$1.5<z<3$; ][]{Brandt01, Reddy04}, and distant LBGs \citep[$3<z<6$;][]{Brandt01, Laird06, L05, Cowie11}. These stacking studies highlight the use of X-rays from normal galaxies to study the star formation history of the Universe. For example, \cite{Symeonidis11} correlate X-ray emission with infrared (IR) emission for galaxies at $0<z<2$ and show that X-ray emission can be calibrated as a useful star formation indicator to measure the global SFR density. 

In this paper, we perform stacking analyses of LBGs, using the deepest X-ray observations to date \citepalias[4 Ms Chandra observations of CDF-S;][]{X11}, adding newly discovered LBGs and covering the broad redshift range, $1.5 \lesssim z \lesssim 8$ (\ie corresponding to when the age of the Universe was 4.5--0.6 Gyrs). \cite{Cowie11} have completed a study of X-ray emission in this sample to find that X-ray emission from $z=0-8$ LBGs is consistent with normal galaxy populations, provided that a reasonable range of dust attenuation factors ($\sim3-5$) are present. However, \cite{Cowie11} did not attempt to correct for dust attenuation and search for low levels of evolution in the X-ray emission. 
For the first time, we measure an empirical law that relates X-ray luminosity, SFR and redshift ($z$) and compare our results with X-ray binary evolution models (see \citetalias{F12}).

Our paper is organized as follows: we introduce the samples and describe our stacking analysis techniques in Section \ref{sec:data}; we discuss the individually detected X-ray sources in Section \ref{sec:dets} and the stacking results for $z\lesssim4$ and $z\gtrsim 5$ in Sections \ref{sec:Xray-sf} -- \ref{sec:Xray-BH}; finally, we offer our conclusions in Section \ref{sec:end}. 
The Galactic column densities are $8.8\times10^{19}$ cm$^{-2}$ for the E-CDF-S \citep{Stark92}. All of the X-ray fluxes and luminosities quoted throughout this paper have been corrected for Galactic absorption. We assume the standard $\Lambda$CDM cosmology: $\Omega_{M}$=0.3, $\Omega_\Lambda$=0.7, and H$_0$=70 km s$^{-1}$ Mpc$^{-1}$, and the initial mass function (IMF) described by \cite{Kroupa}.   

\section{Data and Analysis}\label{sec:data}
\begin{figure}
\begin{center}
\hspace{-0.1in}
  \includegraphics[width=3.5in]{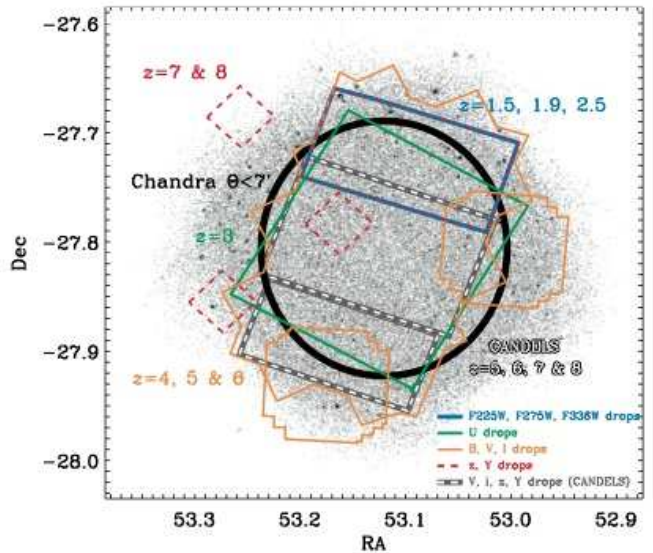}
    \end{center}
  \caption{We show the areal coverage of the different LBG samples, overlaid on the 4~Ms \Chandra full band (0.5--8 keV) image of CDF-S. The F225W- ($z=1.5$),  F275W- ($z=1.9$) and F336W- ($z=$2.5) dropouts fall within the blue region; U-dropouts ($z=3$) lie within the green region;  B$_{435}$- ($z=3.8$), V$_{606}$- ($z=5.0$), and i$_{775}$- ($z=5.9$) dropouts are located within the orange contour; z$_{850}$- ($z=6.9$) and Y$_{105}$- ($z=8.0$) dropouts are bounded by the red dashed lines. Additional V$_{606}$-, i$_{775}$-, z$_{850}$-, and Y$_{105}$-dropouts reside in the CANDELS fields (shown in gray and white dashed lines). The thick black line shows the \Chandra area with off-axis angle $<$7\arcmin. }
   \label{fig:cdfs-field}
\vspace{0.05in}
   \end{figure}

\begin{deluxetable*}{cccccccc}
\tabletypesize{\scriptsize}
\setlength{\tabcolsep}{0.05in}
\tablecolumns{8} 
\tablewidth{7in} 
\tablecaption{Lyman Break Galaxy Samples for Deep Chandra X-ray Study} 
\tablehead{
& & & \colhead{Area} &  & \colhead{$\langle \rm m_{UV} \rangle$ \tablenotemark{a}}&
 \colhead{$\langle \rm SFR_{UV, corr} \rangle$ \tablenotemark{b}} & \colhead{reference}\\
\colhead{$z$} & \colhead{dropout band} & \colhead{instrument}   &  \colhead{(arcmin$^{2}$)} & \colhead{N$_{\rm LBGs}$} & \colhead{(mag)}&
 \colhead{(\msunyr)} & \\
}
\startdata

1.5$\pm 0.4$ & F225W (2372\AA) & \HST/WFC3 & 50 & 48 & 24.5 & 13 & \cite{Oesch2010}\tablenotemark{$\dagger$}\\
1.9$\pm 0.4$ & F275W  (2710\AA) & \HST/WFC3 & 50 & 91 & 24.8 & 30 & \cite{Oesch2010}\tablenotemark{$\dagger$}\\
2.5$\pm 0.6$ & F336W (3355\AA) & \HST/WFC3 & 50 & 359 & 25.4 & 17 & \cite{Oesch2010}\tablenotemark{$\dagger$}\\
3.0$\pm 0.2$ & U (3570\AA) & CTIO + \HST/ACS & 160 & 361 &  25.3 &  33 & \cite{Lee06}\tablenotemark{$\dagger$}\\
3.8$\pm 0.3$ & B$_{435}$ (4317\AA) & \HST/ACS & 196 &2098 & 26.5 & 8 & \cite{Bouwens07}\\
5.0$\pm 0.3$ & V$_{606}$ (5918\AA) & \HST/ACS & 196 &445 & 26.5 & 10 & \cite{Bouwens07}\\
  &   & \HST/WFC3 (CANDELS) & 105 & 768 (700)\tablenotemark{*} & 26.8 & 10 & \cite{Bouwens2012}\\
5.9$\pm 0.3$  & i$_{775}$ (7693\AA) & \HST/ACS & 196 &181 & 27.0 & 5 & \cite{Bouwens07}\\
  &   & \HST/WFC3 (CANDELS) & 105 &  218 (208)\tablenotemark{*} & 27.0 & 8 & \cite{Bouwens2012}\\
6.8$\pm 0.4$  &  z$_{850}$ (9055\AA) & \HST/ACS & 150& 73 & 27.9 & 4 & \cite{Bouwens11}\\
  &   & \HST/WFC3 (CANDELS) & 105 &41 (31)\tablenotemark{*} & 26.7 & 11 & \cite{Bouwens2012}\\
8.0$\pm 0.5$  & Y$_{105}$ (1.055$\mu m$) & \HST/WFC3 & 150 & 60 & 28.0 & 2 & \cite{Bouwens11}\\
  &   & \HST/WFC3 (CANDELS) & 105 & 24 (22)\tablenotemark{*} & 27.0 & 5 & \cite{Bouwens2012}\\
\tablenotetext{a}{Mean apparent rest-frame UV magnitude. }
\tablenotetext{b}{Mean dust-corrected UV SFR. }
\tablenotetext{$\dagger$}{catalog of sources was acquired via private communication.}
\tablenotetext{*}{The total number of LBGs in the dropout sample, with the number of unique (not included in ACS sample) LBGs in parentheses.}
\enddata
\label{tab:LBGs} 
\end{deluxetable*}

\begin{figure}[h]
\begin{center}
  \includegraphics[width=3.5in]{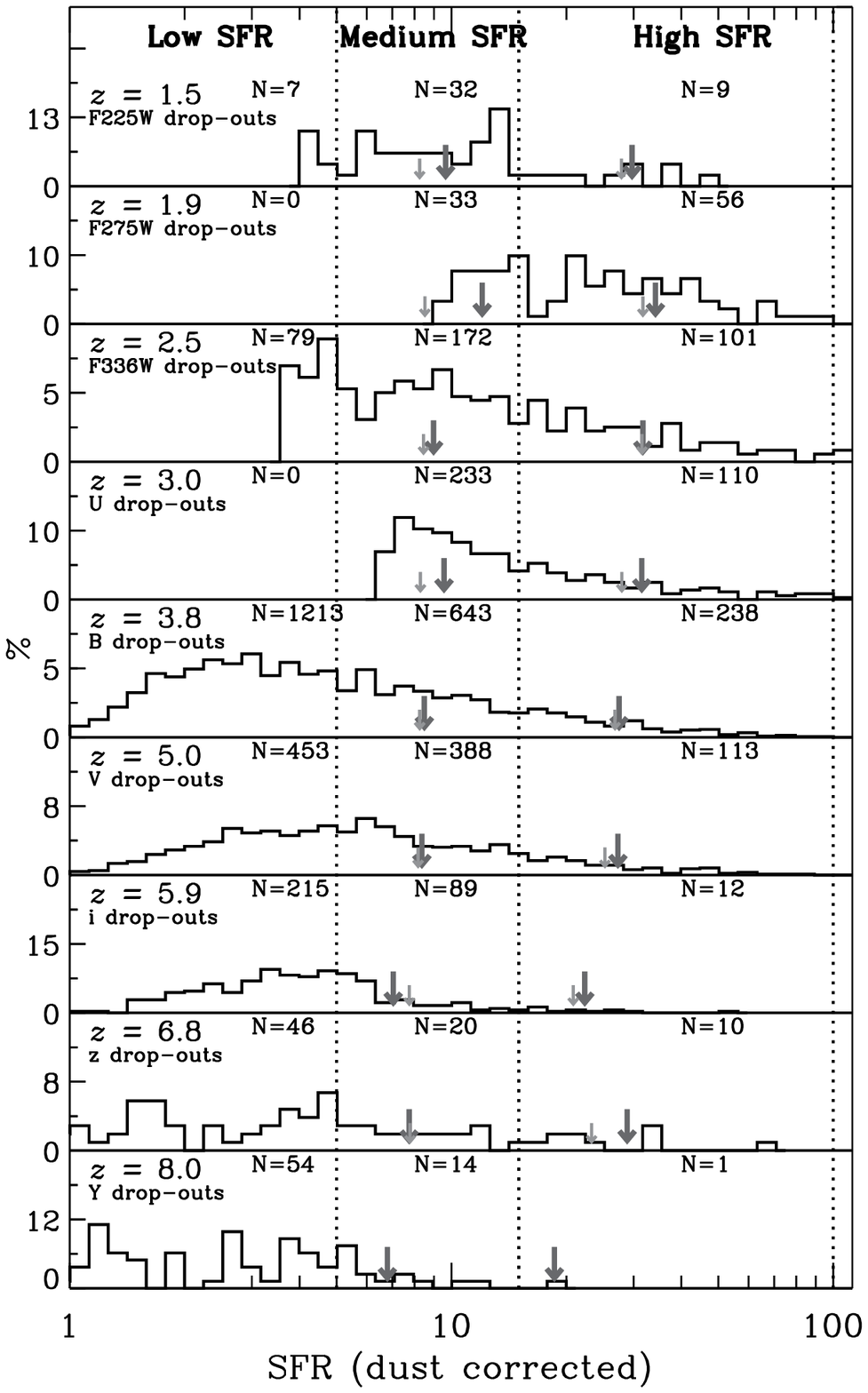}
    \end{center}
  \caption{We show the dust corrected UV-derived SFR distributions of the LBG samples used in this study. The dotted lines mark our SFR cuts for two separate stacking bins: SFR/(M$_\odot$ yr$^{-1}$)$=5-15$ and SFR/(M$_\odot$ yr$^{-1}$)$=15-100$ with the numbers of galaxies that fall in each bin labeled. The arrows mark the observed mean (dark gray) for the different SFR bins compared to the expected mean (light gray; based on the UVLF). The lowest SFR bins do not yield any stacked detections and suffer from incompleteness; therefore, they are not discussed further.}
   \label{fig:SFRhist}
\vspace{0.05in}
   \end{figure}
   
For our analysis of the X-ray properties of LBGs, we use the publicly available 4~Ms CDF-S catalog and data products from \citetalias{X11}\footnote{Data products are available online at \url{http://www2.astro.psu.edu/\~{ }niel/cdfs/cdfs-chandra.html}}. 

We use $1.5 \lesssim z \lesssim 8$ LBG samples compiled from several catalogs (see Table \ref{tab:LBGs}): $z\approx1.5, 1.9$ and 2.5 from \cite{Oesch2010}, $z\approx3$ from \citet{Lee06}, $z\approx3.8, 5.0$ and 5.9 from \cite{Bouwens07}, and $z\approx6.8$ and 8.0 from \cite{Bouwens11}. We include additional $z\approx5-8$ LBGs from the Multi-Cycle Treasury Program Cosmic Assembly Near-Infrared Deep Extragalactic Legacy Survey (CANDELS) CDF-S field \citep{Bouwens2012, Oesch2012b}. We refer readers to these papers for details about the LBG selection, however the basic technique for selecting these galaxies involves using photometric color selection to isolate galaxies with relatively flat spectra that drop out of the bluest filter, while avoiding the expected color tracks of nearby interlopers (stars, low-redshift elliptical galaxies). 

The areal coverages of the LBG samples are shown in Figure \ref{fig:cdfs-field} on top of the \Chandra 0.5--8 keV full-band image. The blue region marks the LBGs selected from the Hubble Space Telescope ({\it HST}) Wide Field Camera 3 (WFC3) Early Release Science data, \ie galaxies dropping out of the F225W ($z\approx1.5$; 48 galaxies), F275W($z\approx1.9$; 91 galaxies), and F336W($z\approx2.5$; 359 galaxies) filters. The green region contains the 361 $z\approx3.0$ LBG sample, selected as U-dropouts, based on ground-based U-band observations taken on the 4m telescope at Cerro Tololo Inter-American Observatory (CTIO) of the Great Observatories Origins Deep Survey-South (GOODS-S) field. The orange contour bounds the samples of 2098 B$_{435}$-($z\approx3.8$), 445 V$_{606}$- ($z\approx5.0$) and  181 i$_{775}$- ($z\approx5.9$) dropouts selected using the Advanced Camera for Surveys (ACS) instrument aboard {\it HST}, which includes the GOODS-S field and two additional regions from parallel fields for NICMOS Ultra-deep Field (UDF; PI: Roger Thompson; \citealt{Thompson}). The red dashed regions mark the locations of 73 $z\approx7$ and 60 $z\approx8$ LBGs, selected from dropping out of the z$_{850}$- and Y$_{105}$ filters using \HST ACS and WFC3 infrared data. The CANDELS fields, shown with gray and white dashed lines (CANDELS-Deep in upper region; CANDELS-Wide field in lower), add 700 V$_{606}$-, 208 i$_{775}$-, 31 z$_{850}$- and 22 Y$_{105}$-dropouts. This information is summarized in Table \ref{tab:LBGs}. 

Contamination from various sources (\eg low-mass stars, AGN, low redshift galaxies, transient sources, photometric scatter and spurious sources) in the LBG sample is expected to range from 3\% to 10\%, and is discussed in more detail in \citet{Bouwens07} and \citet{Bouwens11}

 \subsection{LBG samples: Determining UV-based Dust-corrected SFRs}\label{sec:SFR}  
\begin{deluxetable*}{clllclclc}
\tabletypesize{\scriptsize}
\setlength{\tabcolsep}{0.05in}
\tablecolumns{8} 
\tablewidth{6.5in} 
\tablecaption{Stellar Mass Properties for $z<$ 6 Lyman Break Galaxies} 
\tablehead{
&  & &  \colhead{$\langle\rm \log M_* \rangle$} &  \colhead{Med ($\log$ M$_*$)\tablenotemark{*}} & \colhead{$\langle\rm \log SFR \rangle$} &  \colhead{Med ($\log$ SFR)\tablenotemark{*}} &  \colhead{$\langle \rm \log sSFR \rangle$} &  \colhead{Med ($\log$ sSFR)\tablenotemark{*}} \\
 \colhead{$z$} &  \colhead{N$_{\rm match}$\tablenotemark{a}} &  \colhead{N$_{\rm tot}$\tablenotemark{b}} & \colhead{(\msun)} & \colhead{(\msun)} &  \colhead{(\msunyr)} & \colhead{(\msunyr)} & \colhead{(yr$^{-1}$)} & \colhead{(yr$^{-1}$)} 
}
\startdata
\cutinhead{All SFRs}  1.5   &    42   &    47   &   9.6  $\pm{0.6}$ &   9.5   &   1.0  $\pm{0.3}$ &   1.0   &  -8.6  $\pm{0.6}$ &  -8.4   \\
  1.9   &    59   &    91   &   9.5  $\pm{0.5}$ &   9.4   &   1.4  $\pm{0.3}$ &   1.4   &  -8.1  $\pm{0.5}$ &  -8.1   \\
  2.5   &   127   &   358   &   9.4  $\pm{0.7}$ &   9.4   &   1.2  $\pm{0.4}$ &   1.2   &  -8.2  $\pm{0.6}$ &  -8.2   \\
  3.0   &   156   &   360   &   9.5  $\pm{0.6}$ &   9.5   &   1.2  $\pm{0.3}$ &   1.1   &  -8.3  $\pm{0.5}$ &  -8.3   \\
  3.8   &   982   &  1847   &   8.9  $\pm{0.6}$ &   8.9   &   0.8  $\pm{0.4}$ &   0.7   &  -8.1  $\pm{0.5}$ &  -8.1   \\
  5.0   &   414   &   874   &   8.9  $\pm{0.8}$ &   8.9   &   0.8  $\pm{0.3}$ &   0.8   &  -8.1  $\pm{0.7}$ &  -8.1   \\
  5.9   &   136   &   292   &   8.9  $\pm{0.9}$ &   9.0   &   0.7  $\pm{0.3}$ &   0.7   &  -8.2  $\pm{0.8}$ &  -8.3   \\
  6.8   &     4   &    65   &   9.5  $\pm{0.6}$ &   9.9   &   1.5  $\pm{0.2}$ &   1.5   &  -8.0  $\pm{0.4}$ &  -8.0   \\
\cutinhead{medium SFRs : SFR$=5-15$ \msunyr}
  1.5   &    28   &    31   &   9.6  $\pm{0.6}$ &   9.6   &   1.0  $\pm{0.1}$ &   1.0   &  -8.7  $\pm{0.7}$ &  -8.5   \\
  1.9   &    17   &    28   &   9.6  $\pm{0.8}$ &   9.3   &   1.1  $\pm{0.1}$ &   1.1   &  -8.5  $\pm{0.8}$ &  -8.3   \\
  2.5   &    49   &   167   &   9.2  $\pm{0.7}$ &   9.2   &   0.9  $\pm{0.1}$ &   0.9   &  -8.3  $\pm{0.7}$ &  -8.2   \\
  3.0   &    89   &   224   &   9.3  $\pm{0.5}$ &   9.3   &   1.0  $\pm{0.1}$ &   1.0   &  -8.3  $\pm{0.5}$ &  -8.3   \\
  3.8   &   357   &   580   &   9.0  $\pm{0.5}$ &   9.1   &   0.9  $\pm{0.1}$ &   0.9   &  -8.1  $\pm{0.5}$ &  -8.1   \\
  5.0   &   196   &   348   &   9.0  $\pm{0.8}$ &   9.0   &   0.9  $\pm{0.1}$ &   0.9   &  -8.1  $\pm{0.7}$ &  -8.1   \\
  5.9   &    53   &    80   &   8.9  $\pm{0.9}$ &   9.0   &   0.8  $\pm{0.1}$ &   0.8   &  -8.0  $\pm{0.8}$ &  -8.0   \\
\cutinhead{high SFRs: SFRs $=15-100$\msunyr}
  1.5   &     8   &     9   &   9.8  $\pm{0.5}$ &   9.8   &   1.5  $\pm{0.1}$ &   1.5   &  -8.3  $\pm{0.6}$ &  -8.1   \\
  1.9   &    40   &    61   &   9.5  $\pm{0.4}$ &   9.5   &   1.5  $\pm{0.2}$ &   1.5   &  -8.0  $\pm{0.3}$ &  -8.0   \\
  2.5   &    56   &   105   &   9.7  $\pm{0.5}$ &   9.7   &   1.5  $\pm{0.2}$ &   1.5   &  -8.2  $\pm{0.5}$ &  -8.1   \\
  3.0   &    64   &   119   &   9.8  $\pm{0.5}$ &   9.7   &   1.5  $\pm{0.2}$ &   1.4   &  -8.3  $\pm{0.5}$ &  -8.2   \\
  3.8   &   156   &   231   &   9.5  $\pm{0.5}$ &   9.5   &   1.4  $\pm{0.2}$ &   1.4   &  -8.1  $\pm{0.5}$ &  -8.1   \\
  5.0   &    57   &   112   &   9.5  $\pm{0.7}$ &   9.5   &   1.4  $\pm{0.2}$ &   1.3   &  -8.1  $\pm{0.7}$ &  -8.2   \\
  5.9   &    10   &    14   &   9.9  $\pm{0.4}$ &  10.0   &   1.3  $\pm{0.2}$ &   1.4   &  -8.6  $\pm{0.3}$ &  -8.5   \\
  6.8   &     4   &     9   &   9.5  $\pm{0.6}$ &   9.9   &   1.5  $\pm{0.2}$ &   1.5   &  -8.0  $\pm{0.4}$ &  -8.0   \\

\tablenotetext{*}{Median values}
\tablenotetext{a}{The number of matches with \citet{Xue12} with $|\Delta z|<0.5$}
\tablenotetext{b}{The total number of LBGs within \citet{Xue12} footprint.}
\enddata
\label{tab:mass} 
\end{deluxetable*}
In order to study the relationship between X-ray luminosity and SFRs in LBGs, we estimate their SFRs based on rest-frame UV properties. We use the relation given in Equation 1 of \cite{Bell05} to convert rest-frame UV luminosities into UV-derived SFRs, given by the right-side of this equation: 
\begin{equation}
\rm SFR~(M_\odot yr^{-1}) = 9.8\times10^{-11} (\rm L_{\rm IR} + 2.2 \rm L_{\rm UV})/L_\odot  \label{eqn:sfr}
\end{equation}
where L$_{\rm UV}$ is the UV luminosity ($=\nu{\rm l}_{\nu,2800}$, where $\nu$ is the frequency, and $l_{\nu, 2800}$ is the monochromatic luminosity measured at rest-frame 2800\AA) and L$_\odot$ is solar luminosity (L$_\odot=3.84 \times 10^{33}$ erg s$^{-1}$). Based on the available LBG catalogs, we note that the given UV magnitudes correspond to rest-frame 1500\AA~for some galaxy samples ($z\approx$1.5, 1.9, and 2.5) and to 2800\AA~for the other samples. Since the LBG galaxy spectrum (L$_{\nu}$) is expected to be very flat in this spectral region there are negligible corrections to the SFR based on these differences \citep{Kennicutt}. 

We derive dust corrected SFRs following the prescription given in \cite{Bouwens11dust}. Using the measured UV-continuum slopes ($\beta$) in $z\approx2-7$ LBGs \citep{Bouwens09, Bouwens11dust}, \cite{Bouwens11dust} provide a conversion of the observed UV absolute magnitude and redshift into an estimated $\beta$. We apply the relation between $\beta$ and FUV dust attenuation from \cite{Meurer99} to estimate the extinction factor for each galaxy, which also takes into account the typical scatter of $\beta$ values ($\sim0.36$) at an observed UV magnitude and redshift. These extinction factors range from 3--6 for the $z<3$ LBGs and are $<$2 at $z>3$, with typical uncertainties of $\sim 0.3$ (see Table \ref{tab:radioSFR}). As an additional check on the dust-corrected SFRs, we compare the average dust-corrected UV SFR with radio- and UV$+$IR-derived SFRs for each redshift sample in Section \ref{sec:radio}.

We show the SFR distributions of LBG samples in Figure \ref{fig:SFRhist}. We divide our LBG samples into separate SFR bins (marked by the dotted lines): SFR/(M$_\odot$ yr$^{-1}$)$=1-5$ (low SFR), SFR/(M$_\odot$ yr$^{-1}$)$=5-15$ (medium SFR) and SFR/(M$_\odot$ yr$^{-1}$)$=15-100$ (high SFR) with the numbers of galaxies that fall in each bin labeled. This choice of SFR binning was selected to match the SFR binning at low redshifts (\citetalias{L08}; \citealt{Lehmer2010}) for direct comparison. The dark gray arrows mark the mean for each SFR bin and the light gray marks the expect mean, calculated using the UV luminosity function (UVLF; the form is described by the parameters given in the papers listed in Table \ref{tab:LBGs}  for each LBG sample) to predict the number of galaxies beyond the observed limit. The $z\lesssim3$ medium SFR samples appear to suffer from incompleteness, in that the observed mean SFRs are overestimated compared to the true mean for LBGs. We discuss the implications of this incompleteness on our results in Section \ref{sec:Xray-sf} and \ref{sec:Xray-XLD}.

To compare stellar masses, M$_*$, for these LBGs, we match to the \citet{Xue12} catalog, which uses the available (12 bands) photometric data to estimate stellar masses. We only consider matches where the redshifts (between our sample and the \citealt{Xue12} catalog) agree within $|\Delta z| <0.5$. The number of matches decreases for fainter and higher redshift objects. Only 4 of the $z>6$ LBGs were matched to sources in the \citet{Xue12} catalog; since the \citet{Xue12} catalog has a 5-$\sigma$ z$_{850}$-band limiting magnitude of 28.1, it is likely that most of our highest redshift LBGs are undetected in this filter. The \citet{Xue12} catalog gives the most likely redshift, but it is possible that the second most likely redshift would match the LBG selection redshift (\eg see Section \ref{sec:dets} about XID$=$371). While the $|\Delta z| <0.5$ criterion diminishes the sample size significantly, the average and median M$_*$ offer valuable information about the samples: $\langle\log$M$_*\rangle= 8.9-9.6$ for the full sample; $8.9-9.6$ for the medium SFR sample; and $9.5-9.8$ for the high SFR sample. The specific SFRs (sSFRs$\equiv$SFR/M$_*$) are $\langle \rm \log sSFR\rangle=-8.7$ to $-8.0$. The median and mean values for M$_*$, SFR, and sSFR are given for each sub-sample (redshift and SFR range) in Table \ref{tab:mass}. Given the high sSFRs for these galaxies, the stellar populations in these galaxies are likely relatively young, exhibiting high present to past SFRs.

\begin{deluxetable*}{lllllcccccl}
\tabletypesize{\scriptsize}
\setlength{\tabcolsep}{0.05in}
\tablecolumns{11} 
\tablewidth{7in} 
\tablecaption{Individually Detected X-ray Sources Associated with Lyman Break Galaxies} 
\tablehead{
 &  & & &  &   \colhead{log L$_{UV}$\tablenotemark{a}} & \colhead{SFR\tablenotemark{b}} & \colhead{0.5--8keV\tablenotemark{c}} & \colhead{log L$_{0.5-8}$\tablenotemark{d} } &  &  \\
\colhead{XID} & \colhead{RA} & \colhead{Dec}  & \colhead{$z$} & \colhead{$z_{\rm Xue}$} & \colhead{(erg s$^{-1}$)} & \colhead{(\msunyr)} & (counts) & \colhead{(erg s$^{-1}$)}  & \colhead{$\Gamma$} &  \colhead{Notes\tablenotemark{e} }
}
\startdata
  344   &   53.10485   &  -27.70521   &    1.5   &      1.6 $^{s}$   &  43.91   &      6.3   &       1811.3$^{+50.0}_{-48.7}$   &     43.79 $\pm0.01$   &        2.0  $\pm{0.1}$                            &   L05a, R   \\
  405   &   53.12284   &  -27.72279   &    1.9   &      1.6 $^{p}$   &  43.97   &     13.4   &         59.0$^{+15.1}_{-13.9}$   &     42.93 $\pm0.11$   & $<$    0.1                            &     R   \\
  308   &   53.09392   &  -27.76774   &    1.9   &      1.7 $^{s}$   &  43.95   &     12.3   &        674.5$^{+30.2}_{-29.0}$   &     43.92 $\pm0.02$   &        0.2   $\pm{0.1}$                         &       L05a   \\
  326   &   53.10082   &  -27.71601   &    2.5   &      2.3 $^{s}$   &  44.58   &     54.9   &        234.4$^{+21.7}_{-20.5}$   &     43.46 $\pm0.04$   &        1.6     $\pm{0.2}$    &          R   \\
  137   &   53.03334   &  -27.78257   &    3.0   &      2.6 $^{s}$   &  44.36   &     18.9   &       1249.4$^{+40.4}_{-39.2}$   &     44.45 $\pm0.01$   &        1.3  $\pm{0.1}$   &          R   \\
  490   &   53.14880   &  -27.82112   &    3.0   &      2.6 $^{s}$   &  44.31   &     16.4   &         84.6$^{+12.1}_{-10.9}$   &     43.43 $\pm0.06$   &        0.4     $\pm{0.2}$    &          R   \\
  386\tablenotemark{*}    &   53.11792   &  -27.73432   &    3.0   &      3.6 $^{p}$   &  44.38   &     20.1   &         37.4$^{+11.7}_{-10.6}$   &     42.99 $\pm0.13$   & $>$    1.0                            &          R   \\
  254   &   53.07600   &  -27.87816   &    3.0   &      2.8 $^{s}$   &  44.67   &     44.1   &        246.2$^{+20.6}_{-19.4}$   &     43.65 $\pm0.04$   &        1.8     $\pm{0.2}$    &  L05a, L05   \\
  563   &   53.17439   &  -27.86735   &    3.0   &      3.6 $^{s}$   &  45.43   &    365.0   &       1389.1$^{+42.6}_{-41.4}$   &     44.51 $\pm0.01$   &        1.4   $\pm{0.1}$ &       L05a   \\
  573\tablenotemark{*}   &   53.17848   &  -27.78403   &    3.0   &      3.2 $^{s}$   &  44.31   &     16.4   &        685.1$^{+30.3}_{-29.1}$   &     44.14 $\pm0.02$   &        1.4  $\pm{0.1}$ & L05a, L05   \\
  577   &   53.18015   &  -27.82060   &    3.0   &      1.9 $^{s}$   &  45.41   &    342.6   &       2437.0$^{+54.5}_{-53.3}$   &     44.67 $\pm0.01$   &        1.6 $\pm{0.1}$   & L05a, L05   \\
  588   &   53.18464   &  -27.88092   &    3.0   &      3.5 $^{s}$   &  44.80   &     64.0   &        322.4$^{+24.3}_{-23.1}$   &     43.79 $\pm0.03$   &        1.8     $\pm{0.2}$    & L05a, L05   \\
  388\tablenotemark{*}    &   53.11858   &  -27.88478   &    3.8   &      3.0 $^{p}$   &  44.12   &      7.2   &         41.3$^{+12.7}_{-11.4}$   &     43.38 $\pm0.13$   & $<$    0.3                            &         \nodata   \\
  374   &   53.11163   &  -27.86078   &    3.8   &      3.7 $^{p}$   &  44.48   &     19.3   &                      $<$  26.5   & $<$ 43.11             & $<$    0.7                            &         \nodata   \\
  262   &   53.07848   &  -27.85984   &    3.8   &      3.7 $^{s}$   &  44.50   &     20.2   &        214.6$^{+18.9}_{-17.6}$   &     44.19 $\pm0.04$   &        0.1     $\pm{0.1}$    &          R   \\
  150\tablenotemark{*}    &   53.03989   &  -27.79846   &    3.8   &      3.3 $^{p}$   &  44.11   &      6.9   &         38.6$^{+11.9}_{-10.6}$   &     43.32 $\pm0.13$   & $<$    0.5                            &         \nodata   \\
  100\tablenotemark{*}    &   53.01660   &  -27.74484   &    3.8   &      3.9 $^{p}$   &  44.42   &     16.1   &         93.4$^{+19.3}_{-18.2}$   &     43.63 $\pm0.09$   & $>$    1.1                            &      \nodata         \\
  546   &   53.16528   &  -27.81405   &    3.8   &      3.1 $^{s}$   &  44.75   &     40.0   &       1231.0$^{+39.9}_{-38.7}$   &     44.84 $\pm0.01$   &        0.4    $\pm{0.04}$         & L05a, L05, R   \\
  458   &   53.13850   &  -27.82112   &    3.8   &      3.6 $^{p}$   &  44.49   &     19.8   &         27.6$^{+ 8.0}_{- 6.7}$   &     43.26 $\pm0.12$   &        0.2     $\pm{0.04}$    &         \nodata   \\
  371   &   53.11158   &  -27.76777   &    5.0   &      3.1 $^{p}$   &  44.16   &      7.0   &        173.3$^{+16.3}_{-15.0}$   &     44.19 $\pm0.04$   &        0.6     $^{ + 0.2}_{ -0.1}$    &         \nodata      \\
\enddata
\label{tab:det} 
  \tablenotetext{a}{Observed (\ie not dust-corrected) UV luminosity. }
\tablenotetext{b}{Dust-corrected UV-derived SFR. }
\tablenotetext{c}{Counts refer to the aperture-corrected (background-subtracted) net counts for the full-band ($0.5-8$ keV) from \citetalias{X11}. }
\tablenotetext{d}{L$_{0.5-8}$ is the full band luminosity derived using the 0.5--8 keV observed flux, k-corrected by $(1+z)^{\Gamma-2}$, assuming $\Gamma=1.8$ (typical photon index for AGN), and $z$ (not $z_{\rm Xue}$). Upper limits correspond to 2.5-$\sigma$ limits for sources that were detected in only the hard (2--8 keV) band, while undetected in the soft ($0.5-2$ kev) and full bands. }
\tablenotetext{e}{The code L05 refer to X-ray detected sources in \citetalias{L05}, and were consequently left out of their stacking analyses of \Chandra 1~Ms data. L05a refers to the X-ray sources identified in the E-CDF-S catalog \citep{Lehmer2005ECDFS}. R indicates that this source was detected in the VLA radio data with S/N$>$3 (see Section \ref{sec:radio}).}
\tablenotetext{*}{Found in a close pair ($|\Delta{z}|<0.5$ and angular separations of 3--4\arcsec (corresponding to physical separations of 20--30~kpc at $z\approx3-4$)  with another LBG. See discussion in text (Section \ref{sec:dets}).}
\enddata \label{tab:det} 
\end{deluxetable*} 
   
\subsection{LBG samples: individually detected X-ray sources}\label{sec:X-ray}

Since our primary goal is to study the X-ray emission arising from star formation in these high-redshift LBGs, we avoid including AGN in our X-ray stacking analyses. We assume that sources that have been individually detected at these redshifts (and therefore with L$_{\rm X} >10^{42}$ \ergs) must be dominated by emission from AGN. 

\citetalias{X11} have matched the X-ray detected sources to optical, infrared and radio catalogs using probabilistic matching, which estimates the false match statistics by applying a Monte Carlo technique \citep[see also][]{Broos} and accounts for positional uncertainties (which vary across the \Chandra field with off-axis angle). \citetalias{X11} found an offset between the the optical/infrared positions and radio positions (the X-ray positions are fixed to the radio frame): RA$_{\rm X, radio}=\rm RA_{opt, IR}+$0.175\arcsec and Dec$_{\rm X, radio}=\rm Dec_{\rm opt, IR}-$0.284\arcsec. The coordinates (shifted to the radio/X-ray system) for the best matched GOODS-S or GEMS sources are provided, which we match to our LBG positions within 0.5\arcsec. We find 20 X-ray detected LBGs (which are likely AGN) and tabulate their properties in Table \ref{tab:det}. 

In this Table, the redshifts based on the LBG sample selection ($z$) are listed along with the redshifts from the \citetalias{X11} catalog, based on spectroscopic ($s$) or other photometric data ($p$). The effective photon index ($\Gamma$) was determined for all sources with $>$200 full band (0.5--8keV) counts using basic spectral fitting (described in \citetalias{X11}) and assuming an absorbed power law. The last column in this table includes notes about previously detected and studied sources from CDF-S 1 Ms \citepalias{L05}, E-CDF-S catalog \citep[L05a;][]{Lehmer2005ECDFS}, and radio (R) detections based on our VLA data analysis (see Section \ref{sec:radio}). These sources are discussed in more detail in Section \ref{sec:dets}. 
   
\subsection{X-ray stacking of individually undetected LBGs}\label{sec:stack}
While even 4~Ms of exposure time is not sufficient for detecting X-rays from individual star-forming galaxies at $z>1$ \citep[with the possible exception of submillimeter galaxies;][]{Alexander05, Laird2010}, the X-ray emission from large numbers of galaxies can be stacked to provide the average X-ray properties of the stacked sources. We follow the stacking technique outlined in \citet[][hereafter, \citetalias{L08}]{L08} to stack LBGs in bins of SFR (shown in Figure \ref{fig:SFRhist}) and $z$. 

First, we select sources that have no X-ray detections in the catalog from \citetalias{X11} within twice the 99\% encircled energy fraction (EEF) radius (note that this value changes with off-axis angle) in order to avoid contamination from the \Chandra point spread function (PSF) from detected sources. We only include sources with off-axis angles $<7$\arcmin~from the central pointing (thick black circle in Figure \ref{fig:cdfs-field}) because the increasing PSF at large off-axis angles degrades the X-ray sensitivity significantly. These two criteria eliminate 856 and 837 LBGs, respectively, from our original sample of LBGs. 

Recent stacking results of $z>6$ LBGs by \citet{Treister11} have generated discussion about proper background subtraction in order to avoid artificial stacked detections (\eg by setting the clipping threshold of the background pixels too low; see \citealt{Cowie11,  Willott11}). We refer readers to \cite{Lehmer2005ECDFS} and L08 for detailed descriptions about our background subtraction technique using background maps, which do not include any clipping of photons. Briefly, these background maps were created by filling the regions near detected sources (twice the 90\% PSF EEF radius, r$_{90}$) with noise, which is generated based on the probability distribution of counts in the local background (within an annulus with radii between 2 and 4 $\times$ r$_{90}$). 

Throughout the remainder of this paper, X-ray luminosities refer to the rest-frame hard (2--10 keV) energy band. These luminosities are k-corrected from the observed soft band (0.5--2 keV) since this band is closer to the rest-frame 2--10 keV band, and therefore the $k$-correction is smaller and less affected by the assumed spectral shape (\ie photon index, $\Gamma$). For example, at the median redshift of our LBG samples ($z\approx3$) the observed 0.5--2 keV band corresponds to rest-frame 2--8 keV. The $k$-correction can be calculated as follows:
\begin{eqnarray}
k_{corr}& = & \frac{ E_{out, 2}^{(2-\Gamma)}-E_{out, 1}^{(2-\Gamma)}}{E_{in, 2}^{(2-\Gamma)}-E_{in, 1}^{(2-\Gamma)}} (1+ z)^{(\Gamma-2.0)} \label{eqn:kcor}\\
& = & 2.472~(1+ z)^{-0.3}
\end{eqnarray}
using $\Gamma=1.7$, based on the expected photon index for X-ray binaries \citep{Ptak99} 
and in the input and output energies: $E_{out, 1}=$2 keV, $E_{out, 2}=$10 keV, $E_{in, 1}=$0.5 keV, $E_{in, 1}=$2 keV.\footnote{We note that we attempted to stack in the observed 2--8~keV bandpass, however we did not obtain $>2\sigma$ detections in any cases. Ideally, we could use the 2--8~keV/0.5--2~keV stacked band ratios as proxies for the spectral slopes of our stacks.  Unfortunately, the upper limits of the band ratios are not well constrained ($\Gamma_{\rm eff}>$0.3--0.7) for all cases. Therefore, we limit further discussion to results based on stacking the observed soft 0.5--2~keV band.} 

We consider two sources of error in our stacked measurements. The first, $\sigma_{\rm p}$, is Poisson noise and calculated simply as $\sqrt{\rm B}$, where B represents background counts. This is the error that we use to determine the signal to noise. In addition, we also calculate the errors from boot-strapping ($\sigma_{\rm boot}$), which measure how the contribution of individual sources may affect the average value. To determine our boot-strapping errors, we randomly resample the LBGs for each stacking bin a large number of times (5000) and repeat our stacking analysis on these samples. The random resampling will duplicate some values while eliminating others each iteration to statistically quantify the effect of individual sources on the stack \citep[$\sim 37\%$ of the values are replaced by duplicated values; see][]{DasBoot}. The boot-strapped errors refer to the standard deviation of the stacked values from this random resampling. Bootstrap errors include the variation due to background fluctuations (Poisson error), therefore the uncertainties quoted in our analysis are the bootstrap errors.
      
To study the relationship between X-ray luminosity and SFR of $z\lesssim 4$ LBGs, we stack the LBG samples in two SFR bins:  SFR/\msunyr$=5-15$ and SFR/\msunyr $=15-100$.  For the remainder of our study, we require a S/N cut of 2.5-$\sigma$ for a stacked detection (which corresponds to $>$99.38\% confidence limit for one-sided gaussian distributions). Since the low SFR samples do not yield any detections and suffer greatly from incompleteness, we do not include them in our discussion. The results of our stacking analyses are discussed in detail in Section \ref{sec:Xray-sf}. 

\subsection{Contribution of AGN to average X-ray properties }\label{sec:fagn}

\begin{figure}
\begin{center}
  \includegraphics[width=3.5in]{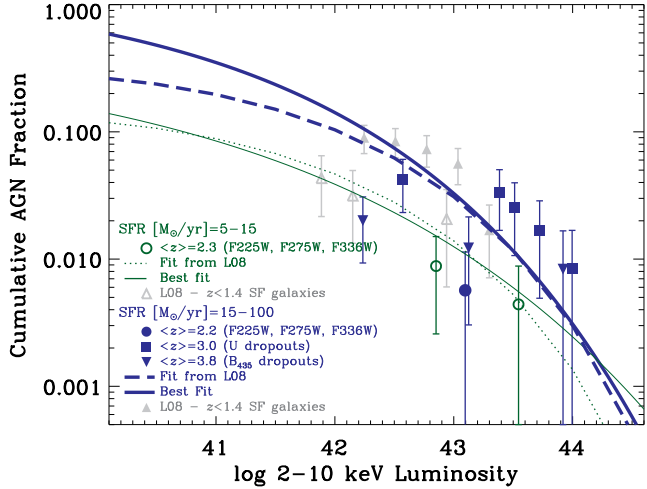}
    \end{center}
  \caption{We show the cumulative AGN fraction vs. 2--10 keV X-ray luminosity for SFR$=$5--15 \msunyr (medium SFR; open green symbols)  and SFR$=$15--100 \msunyr LBGs (high SFR; filled blue symbols) and for $z<1.4$ star-forming galaxies from \citetalias{L08} (gray triangles; filled for high SFR and open for medium SFR). The different symbols describe different LBG samples, as listed in the legend. The solid lines mark the error-weighted exponential best fit to the combined dataset for the high SFR (blue) and medium SFR (green) data; best-fits determined from Equation 8 in \citetalias{L08}, using mean SFRs from the full sample for high SFR and medium SFR galaxies are shown as blue dashed and green dotted lines, respectively.}
   \label{fig:f_AGN}
\vspace{0.05in}
   \end{figure}

The 20 X-ray detected sources in Table \ref{tab:det}, which were classified as AGN following the criteria outlined in Section \ref{sec:X-ray}, are useful for assessing the level of AGN contamination in the stacked samples discussed in the previous section ($\S$ \ref{sec:stack}). 

In Figure \ref{fig:f_AGN}, we show the cumulative AGN fractions as a function of X-ray luminosity for high SFR ($15-100$ \msunyr; filled blue symbols) and medium SFR ($5-15$ \msunyr; open green symbols) LBG samples. The different redshift samples are displayed with different symbols, as described in the legend. 

The data are not sufficient to determine whether there is any variation in the AGN fraction with redshift. Focusing on the high SFR LBG sample (filled blue symbols), where we have some redshift information, there is no obvious redshift trend (\eg the $z<3$ point appears consistent with the $z\approx3.8$ data, while the $z\approx3$ values are higher). We proceed by including high SFR data from $z<1.4$ star-forming galaxies from \citetalias{L08} (filled gray triangles). The fit from \citetalias{L08} (blue dashed line), using the mean SFR from the high SFR sample, compares well with our error-weighted fit (solid blue line) to all the data (\citetalias{L08} and our high SFR sample points). The fit from \citetalias{L08} (green dotted line; open gray triangles show medium SFR galaxies from this study), using the mean SFR from the medium SFR sample, appears consistent with our medium SFR fit (dotted green line) to data (open green circles). 

Our higher redshift data probe a larger volume and extend the relation to higher L$_{\rm X}$. At lower L$_{\rm X}$, the difference between \citetalias{L08} may be from incompleteness at these lower X-ray luminosities. Therefore we estimate the AGN contribution from Equation 8 of \citetalias{L08} and subtract this from the stacked X-ray luminosity as described in \citetalias{L08}. Since the AGN fractions from our analysis appear lower (or equal) to those from \citetalias{L08}, we err on the side of over-compensating for the AGN contribution. Nevertheless, these corrections are smaller than the errors on L$_{\rm X}$ (typically $\sim$20\%). We note that \citetalias{L08} estimated $\sim50-70\%$ AGN contamination in $z\approx3$ LBGs, based on CDF-S 1~Ms analysis. Our increased depth, which improves identification and removal of AGN, may account for the significant decrease in our measured AGN fraction; since low luminosity AGN are more numerous, survey depth does not necessarily scale proportionally with AGN fraction.

\hspace{-0.1in}
 \begin{deluxetable}{cccccc}[h]
\tabletypesize{\scriptsize}
\tablecolumns{6} 
\tablewidth{3.3in}
\tablecaption{Comparing SFRs: UV, Radio, and IR} 
\tablehead{
& &  \multicolumn{4}{c}{$\rm \langle SFR\rangle$ [\msunyrs]} \\ 
\cline{3-6}\\
\colhead{z} & \colhead{N} & \colhead{UV} & 
  \colhead{UV$_{\rm corr}$} & \colhead{radio} 
 & \colhead{UV$+$IR} 
} 
\startdata

  1.5 &    30    &  3  $^{+  2 }_{-1}$  &  11  $^{+ 9 }_{- 5}$  &  $<$   27   &    $<$25       \\
  1.9 &    44    &  3  $^{+  2 }_{-1}$  &  22  $^{+17 }_{-10}$  &  $<$   39   &   $<$ 32           \\
  2.5 &   222    &  3  $^{+  2 }_{-1}$  &  11  $^{+15 }_{-6   }$  &     36$\pm$   10   &   $<$ 33           \\
  3.0 &   311    &  5  $^{+  6 }_{-3}$  &  16  $^{+23 }_{- 9}$  &  $<$   45   &   $<$ 48           \\
  3.8 &  1381    &  2  $^{+  3 }_{-1}$  &   5  $^{+ 7 }_{- 3}$  &  $<$   30   &    $<$ 61         \\
  5.0 &   701    &  3  $^{+  3 }_{-2}$  &   6  $^{+ 8 }_{- 3}$  &  $<$  135   &   $<$ 240           \\
  5.9 &   220    &  3  $^{+  2 }_{-1}$  &   4  $^{+ 3 }_{- 2}$  &  $<$  304   &   $<$ 590           \\
  6.8 &    61    &  2  $^{+  3 }_{-1}$  &   3  $^{+ 6 }_{- 2}$  &  $<$  713   &   $<$2200           \\
  8.0 &    44    &  2  $^{+  3 }_{-1}$  &  2  $^{+  3 }_{-1}$  &  $<$ 1064   &   $<$4800 \\
  \enddata
\tablenotetext{}{The mean UV, dust-corrected UV, radio and UV$+$IR SFRs for full sample of LBGs (only excluding individually-detected X-ray sources and those sources within twice the 99\% EEF radius from other X-ray sources). }
\label{tab:radioSFR} 
\end{deluxetable}


 \subsection{Comparing average SFRs using multiwavelength data: radio and IR SFRs}\label{sec:radio}  
We perform stacking of radio and IR observations to offer independent tests for estimating unattenuated SFRs for the $z<5$ LBG samples. Using VLA (\citealt[in prep]{Millerinprep}; see also \citealt{Miller08}) and PACS {\it Hershel} \citep{Elbaz2011} 
data for the CDF-S, we have stacked all of the LBGs that are not X-ray detected. These measurements provide consistency checks between the radio-derived, UV$+$IR SFR, and the dust corrected UV-derived SFRs.

We generated $50\arcsec \times 50 \arcsec$ cutouts of the radio image around each LBG position, and proceeded to stack these cutouts for each LBG redshift sample. These LBG samples exclude the same galaxies as the X-ray stacking sample (\ie those sources that were individually X-ray detected or near, within twice the 99\% EEF radius, other X-ray detected sources). Incidentally, 8 of the 20 individually X-ray detected LBGs (listed in Table \ref{tab:det}) were also individually detected (with $>3\sigma$) in the VLA observations (marked ``R'' in the last column). Two stacking procedures were used: the first used a straight median for each pixel in the stack, and the second used an average with rejection of the highest and lowest pixel. The latter method is designed to prevent real sources within the $50\arcsec$ fields from seeping into the average and hence output stack image, which is more likely when the number of images to stack is large because of the surface density
of such bright sources. 
These real sources are often extended radio emission from unrelated sources, although a small number of the LBGs themselves were individually detected (7 of the 13 radio-detected LBGs with radio $S/N>5\sigma$ were also X-ray detected and are marked with `R' in the last column of Table \ref{tab:det}) and the rejection prevented them from unduly influencing the resulting stack. For the $z\approx3.8$ and $z\approx5.0$ samples, the rejection was extended from the single highest pixel to the two highest pixels to ameliorate contributions of real detected sources and improve the cosmetics of the stacked image. 

The stacked images were then inspected to evaluate the resulting RMS noise level and any sources recovered by the stacking technique. The results of the stacks produced by the median and average procedures were always consistent in terms of the resulting RMS noise and for the sample that yielded a weak detection. The RMS of the stacks ranged from about 0.2$\mu$Jy to 1.3$\mu$Jy depending on the number of objects in the sample, down from the $\sim$6.4$\mu$Jy noise level typical of the radio image across the CDF-S area.  Stacks made from cutouts using arbitrary positional offsets demonstrated that the RMS noise values in the actual sample stacks were consistent, and these ``blank sky'' stacks produced no false detections. The 1.4GHz radio luminosities were $k-$corrected, assuming a standard power law with a spectral index of 0.7, and converted to SFRs using the relation given in \citet[][see Equation 13; using the Salpeter IMF with mass limits of 0.1 and 100 \msun]{Yun}, but divided by a factor of 1.7 to translate SFRs to the Kroupa IMF. 

Herschel-PACS data covering the GOODS-S \citep[see][for a description of the observations]{Elbaz2011} was downloaded from the HeDaM website\footnote{\ttfamily http://hedam.oamp.fr/GOODS-Herschel/goods-south\_data.php}. For the purposes of this paper we used only the 160$\mu$m data as, for our samples, it probes closer to the peak of the far-infrared SED than the other PACS wavebands and thus provides the most reliable indicator of the IR luminosity. To perform our stacking analyses we extracted 60pixel x 60pixel thumbnails (from both the science and RMS images) centered on the positions of all the X-ray undetected sources that are also undetected at 160$\mu$m. For each of our redshift bins we combined corresponding science thumbnails using a weighted mean, weighting by the $1/\sqrt(RMS)$ value at each pixel position. We use aperture photometry to measure the flux within a circular aperture placed at the centre of each stacked image. The error on the mean was calculated by randomly selecting (with replacement) two-thirds of each redshift sample, performing the our stacking procedure 1000 times, then calculating the standard deviation of the resulting flux distribution. Finally, to determine the significance of each stacked (mean) flux, we compared it against that obtained when stacking the same number of random positions.  The mean flux of the 160$\mu$m-undetected sources was then combined with the mean flux of the detected sources using a weighted mean (weighted according to the number of undetected/detected sources) to give the mean flux of all sources in that bin \citep[see][for more details]{JM12}.
We convert the stacked 160$\mu$m flux into a total IR (8--1000$\mu$m) luminosity using the template spectrum from \citet{CharyElbaz}. Then, using Equation \ref{eqn:sfr}, where L$_{\rm IR}$ is the total IR luminosity, we calculate the UV$+$IR SFRs.  

Table \ref{tab:radioSFR} summarizes the mean observed UV, dust-corrected UV, radio and UV$+$IR SFRs. Only the $z=2.5$ LBG sample was detected in the radio stacking at $\sim3\sigma$; none of the LBG samples were detected in the IR stacking analysis. Within errors, the dust-corrected UV-derived SFRs are consistent with the radio and UV$+$IR SFRs. 


%
 
   \begin{figure}
\begin{center}
  \includegraphics[width=2.5in]{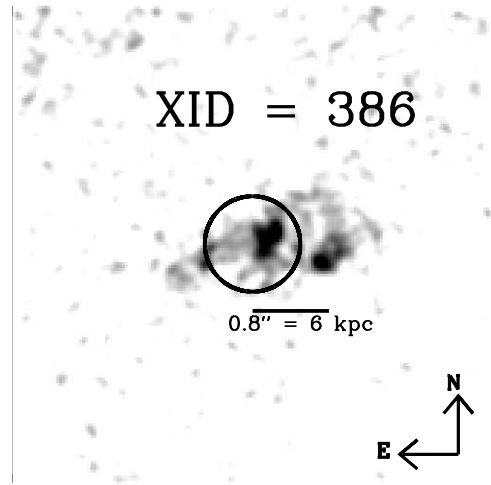}
    \end{center}
  \caption{\HST ACS z$_{850}$ image shows XID$=$386, an X-ray detected LBG at $z\approx3$, apparently interacting with another LBG (also at the same redshift). These LBGs have angular separations of 0.8\arcsec, corresponding to a physical separation of 6~kpc. }
   \label{fig:closepair}
   \end{figure}

\section{Results and Discussion}\label{sec:results}
We divide our discussion of the results into three sections. The first section includes a brief summary of interesting discoveries related to individually detected LBGs from the 4 Ms CDF-S catalog \citepalias{X11}. The other sections relate to our stacking analyses. Based on our stacking results, we find that the $z\lesssim4$ samples provide significant stacked detections, whereas we were only able to measure upper limits for the $z\gtrsim 5$ LBGs. Therefore, we consider it is most meaningful to split the stacking analysis into two parts: studying the evolution of the X-ray properties at $z\lesssim4$ as related to X-ray binaries and star formation, and constraining supermassive black hole growth for $z\approx 3-8$ LBGs.

\begin{figure}
\begin{center}
  \includegraphics[width=1.1in]{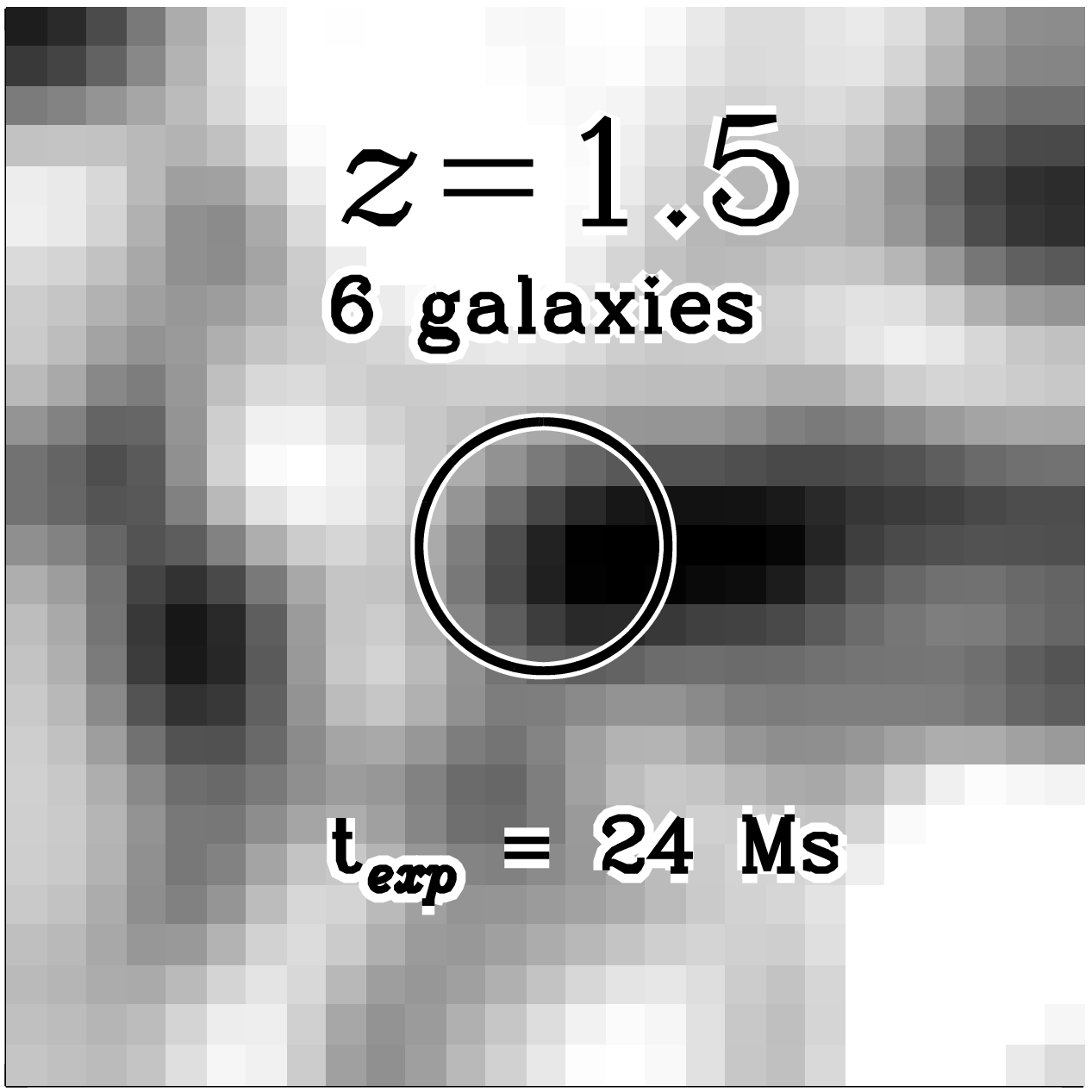}
  \includegraphics[width=1.1in]{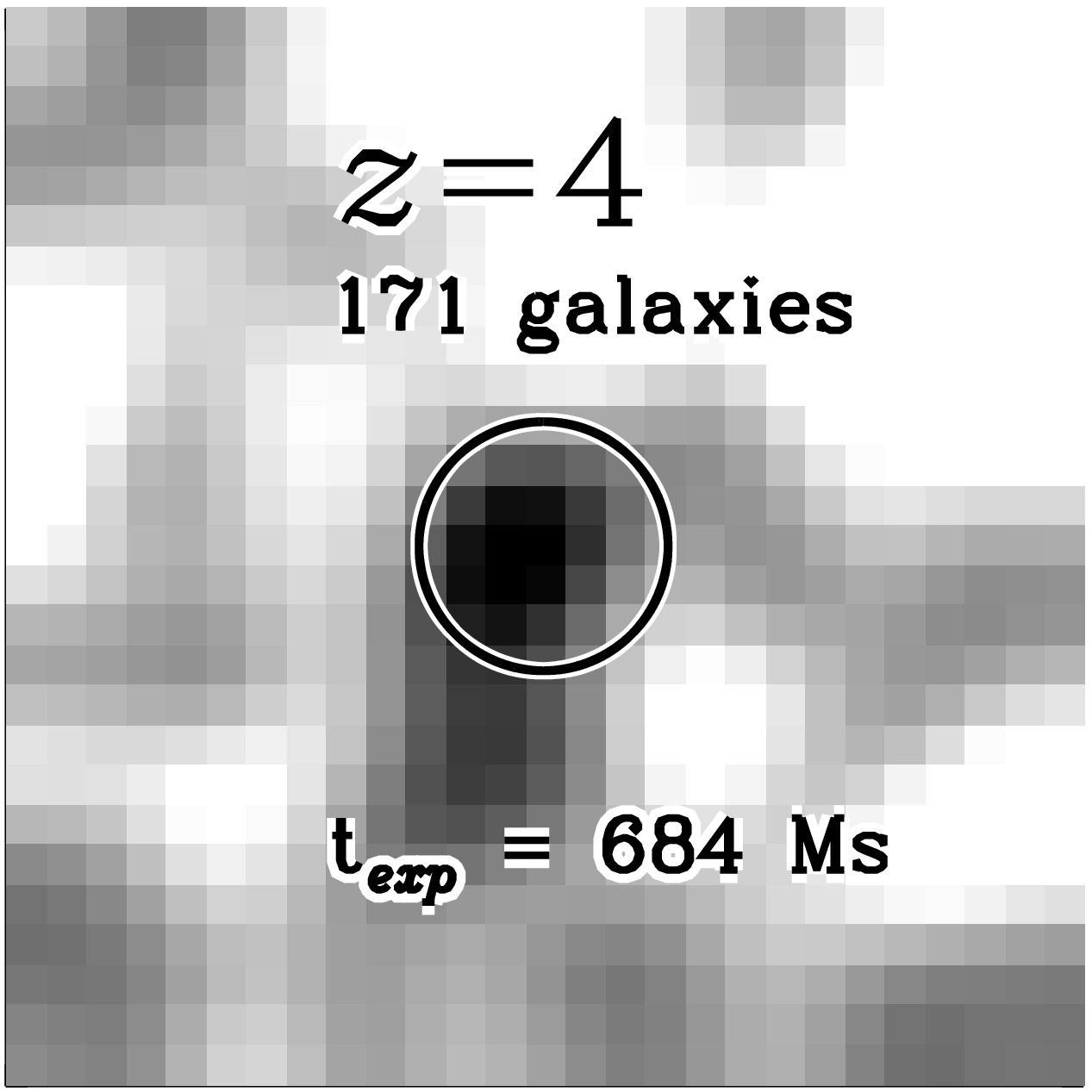}
  \includegraphics[width=1.1in]{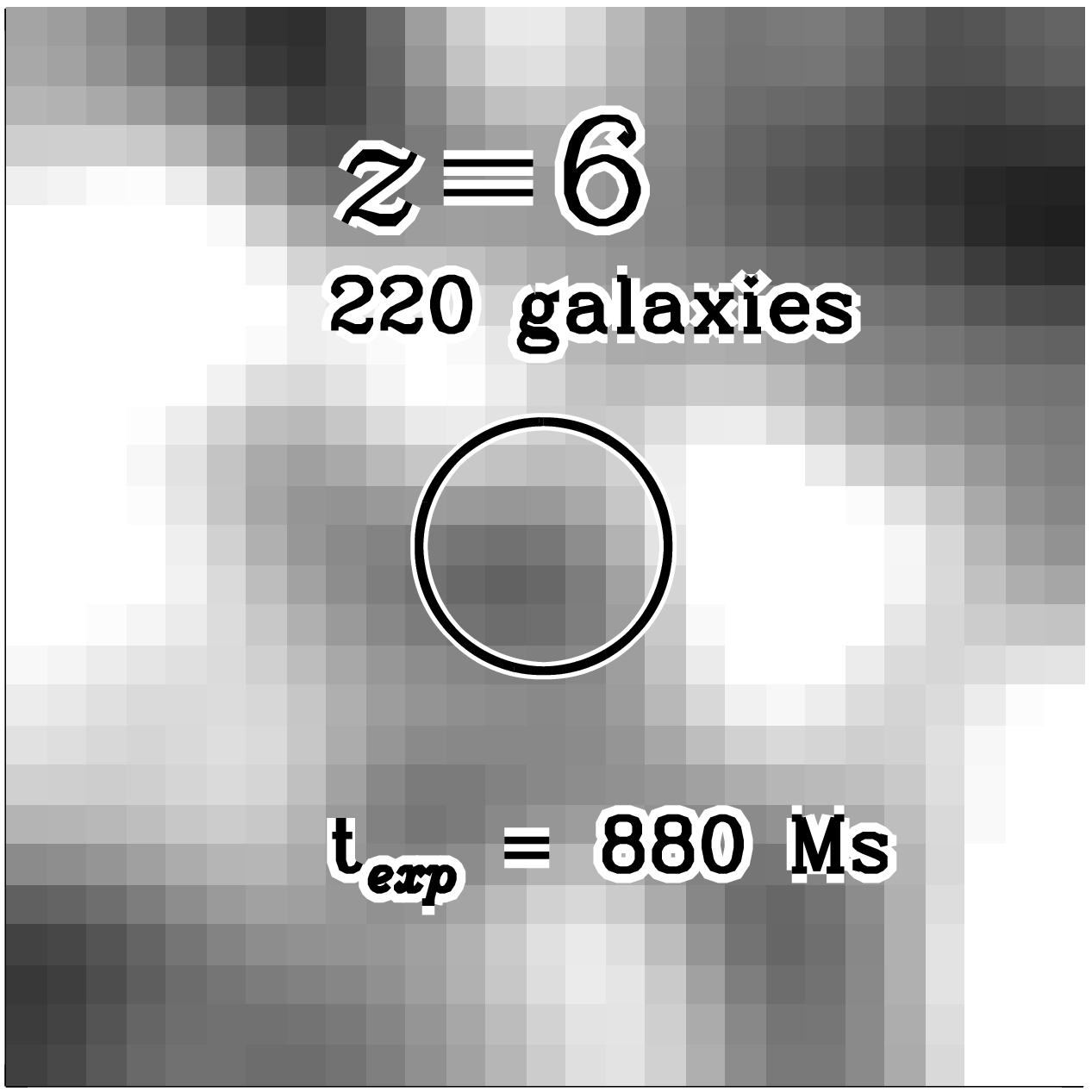}
    \end{center}
  \caption{The observed 0.5--2~keV stacked images for the $z\approx1.5$ and 4 LBGs with SFR$=15-100$ \msunyr (left and middle) and $z=6$ (right) LBG samples are shown. These images have been smoothed by a 3 pixel kernel using gaussian weighting. The $z=1.5$ image shows a 2.6-$\sigma$ detection (near our detection threshold) and contains 6 galaxies (24 Ms exposure), $z=4$ image shows a 4.0-$\sigma$ detection and contains 171 galaxies, corresponding to 684 Ms of exposure. The $z=6$ image contains 220 galaxies ($\sim 880$ Ms total exposure) with S/N$=1.0$. The marked circles have 3\arcsec radii.}
   \label{fig:imstack}
\vspace{0.05in}
   \end{figure} \begin{deluxetable*}{ccccccccc}
\tabletypesize{\scriptsize}
\setlength{\tabcolsep}{0.02in}
\tablecolumns{9} 
\tablewidth{6.5in} 
\tablecaption{Summary of Stacked X-ray/Star formation properties in $z<$4 LBGs} 
\tablehead{
\colhead{$z$} & \colhead{\# of sources} & \colhead{t$_{tot}$\tablenotemark{a}} & \colhead{Net counts\tablenotemark{b}} & \colhead{S/N\tablenotemark{c}} & \colhead{$\log$ L$_{\rm X}$\tablenotemark{d}} & \colhead{$\log$ L$_{\rm UV}$} & \colhead{log SFR\tablenotemark{e}} & \colhead{$\log$ L$_{\rm X}$/SFR}  \\
 &  &  \colhead{(Ms)} & & & \colhead{(erg s$^{-1}$)} & \colhead{(erg s$^{-1}$)}  & \colhead{(M$_{\odot}$ yr$^{-1}$)} & \colhead{(erg s$^{-1}$ [M$_{\odot}$ yr$^{-1}$]$^{-1}$)}  \\
}
\startdata
 \cutinhead{5$<$SFR/(\msunyr)$<$15}
  1.5   &   21   &   84   &      40.3 $\pm$  13.5   &  3.6   &     41.06 $\pm$  0.13   &  44.05   &   1.01 &       40.05 $\pm$  0.13   \\
  1.9   &   15   &   60   & $<$  23.3\tablenotemark{*}               &  2.0   & $<$ 41.14               &  43.94   &   1.08 &   $<$ 40.06               \\
  2.5   &   99   &  396   & $<$  60.3               &  1.5   & $<$ 41.06               &  44.01   &   0.95 &   $<$ 40.11               \\
  3.0   &  191   &  764   &     133.7 $\pm$  35.5   &  4.0   &     41.20 $\pm$  0.10   &  44.11   &   0.98 &       40.21 $\pm$  0.10   \\
  3.8   &  427   & 1708   & $<$ 123.6               &  1.8   & $<$ 41.07               &  44.18   &   0.93 &   $<$ 40.14               \\
\cutinhead{15$<$SFR/(\msunyr)$<$100}
  1.5   &    6   &   24   &      15.3 $\pm$   3.0   &  2.6   &     41.20 $\pm$  0.08   &  44.38   &   1.46 &       39.71 $\pm$  0.08   \\
  1.9   &   28   &  112   &      48.9 $\pm$  14.4   &  3.8   &     41.27 $\pm$  0.11   &  44.22   &   1.48 &       39.75 $\pm$  0.11   \\
  2.5   &   62   &  248   &      81.1 $\pm$  16.8   &  4.3   &     41.41 $\pm$  0.08   &  44.40   &   1.51 &       39.86 $\pm$  0.08   \\
  3.0   &  104   &  416   &      90.0 $\pm$  30.4   &  3.7   &     41.28 $\pm$  0.13   &  44.54   &   1.50 &       39.72 $\pm$  0.13   \\
  3.8   &  171   &  684   &     126.2 $\pm$  32.3   &  4.0   &     41.48 $\pm$  0.10   &  44.60   &   1.44 &       39.98 $\pm$  0.10   \\
    
\tablenotetext{a}{Total exposure time of the stack.}
\tablenotetext{b}{Background-subtracted counts in observed 0.5-2 keV band. Errors include Poisson and bootstrap errors (as described in Section \ref{sec:stack}). }
\tablenotetext{c}{S/N is measured as S/$\sqrt{B}$, where S and B are net and background counts, respectively.}
\tablenotetext{d}{Rest frame 2--10 keV luminosity, $k$-corrected from observed 0.5--2 keV. See text for more details.}
\tablenotetext{e}{Dust-corrected SFR}
\tablenotetext{*}{Upper limits are 2.5-$\sigma$ values.}
\enddata
\label{tab:stackSF} 
\end{deluxetable*} 

\subsection{Individually detected X-ray sources}\label{sec:dets}
We use the probabilistic matches to optical/IR sources from the \citetalias{X11} catalog (see Section \ref{sec:X-ray}) to identify 20 LBGs, with redshifts ranging from $z=1.5-5.0$, that were individually detected in the \Chandra 4~Ms data. The \citetalias{X11} catalog lists photometric or spectroscopic redshifts for all the sources (shown in column 5 in Table \ref{tab:det}). Therefore, the redshifts determined by the LBG dropout technique offer another measurement for comparison. The spectroscopic redshifts match those of the dropout technique reasonably well (8 of the 12 redshifts agree within $|\Delta z|<0.5$). The 4 cases where the spectroscopic redshift differs significantly ($|\Delta z|\geq0.5$) from the dropout redshift include 3 $U$-dropouts ($z\approx3.0$; XIDs$=$563, 577, and 588) and 1 $B_{435}$-dropout ($z\approx3.8$; XID$=$546), and in all these cases the spectroscopic redshift quality is ``secure'' ($\geq 95\%$ confidence levels with multiple spectral features). With the exception of XID=563, all of these were detected in \citetalias{L05} and with the same dropout classification. While the discrepancies in values between the spectroscopic redshifts and the LBG selection seem considerably higher than the expected contamination rate (3--10\%, see Section \ref{sec:data}), the redshift failure rates in these X-ray detected LBGs are not representative of the LBG sample in general. In fact, these 4 AGN are the most UV luminous of our entire sample, and therefore, most likely to contain high-ionization emission lines that are not modeled by the LBG selection. 

The highest redshift LBG sample containing detected sources is $z=5$, with a single X-ray detected source (XID$=$371). While the photometric redshift for this source in the \citetalias{X11} catalog is $z=3.1$, a more recent determination utilizing the CANDELS photometry gives a more likely redshift of $z=4.65^{+0.18}_{-0.41}$ \citep{Xue12}.  

   
We find that 5 of the 20 X-ray detected LBGs appear to be in close pairs (with other LBGs, in the same redshift sample, within 30~kpc; marked with asterisks in Table \ref{tab:det}). For comparison, 512 of the 4419 LBGs ($\sim 12\%$) are in close pairs.  Therefore, based on our crude estimate, X-ray detected LBGs appear to have a higher probability (factor of 2 enhancement) of being in a close pair than other LBGs. However, our analysis does not calculate the real space pair fraction (\ie the width of the redshift bins, listed in Table \ref{tab:LBGs}, correspond to large physical distances) nor takes into account the possibility of chance projections; therefore, we stress that a more detailed statistical analysis is required to measure accurate pair fractions and test the significance of close pair enhancement in X-ray detected LBGs. In one case, XID=386, the other LBG is separated by 0.8\arcsec (at $z=3$, this corresponds to a physical separation of 6~kpc; Figure \ref{fig:closepair}). Inspecting the \HST ACS $z_{850}$ image for this source, we observe two bright peaks, surrounded by diffuse extended emission. Given that the observed sizes of LBGs at $z\sim3$ are typically smaller ($\sim 2-3$ kpc; \citealt{Ferguson04,Trujillo06,Franx08,Mosleh11}), it is unlikely that the LBGs belong to a single galaxy larger than 6~kpc. Rather, we speculate that these two separate galaxies are undergoing a merger. These sources also correspond to a detected radio source of $\sim 8\sigma$. However, none of the other 4 close pairs were radio detections; in these cases, the angular separations are 3--4\arcsec, corresponding to physical separations $>20-30$~kpc at $z=3.0$ and 3.8. While further study of these pairs is beyond the scope of this paper, we note this potentially relevant observation and conjecture that the enhanced star formation and AGN activity could be related to interactions in these LBGs \citep[see also][]{Kocevski}.
   \begin{figure}
\begin{center}
\hspace{-0.4in}
  \includegraphics[width=3.7in]{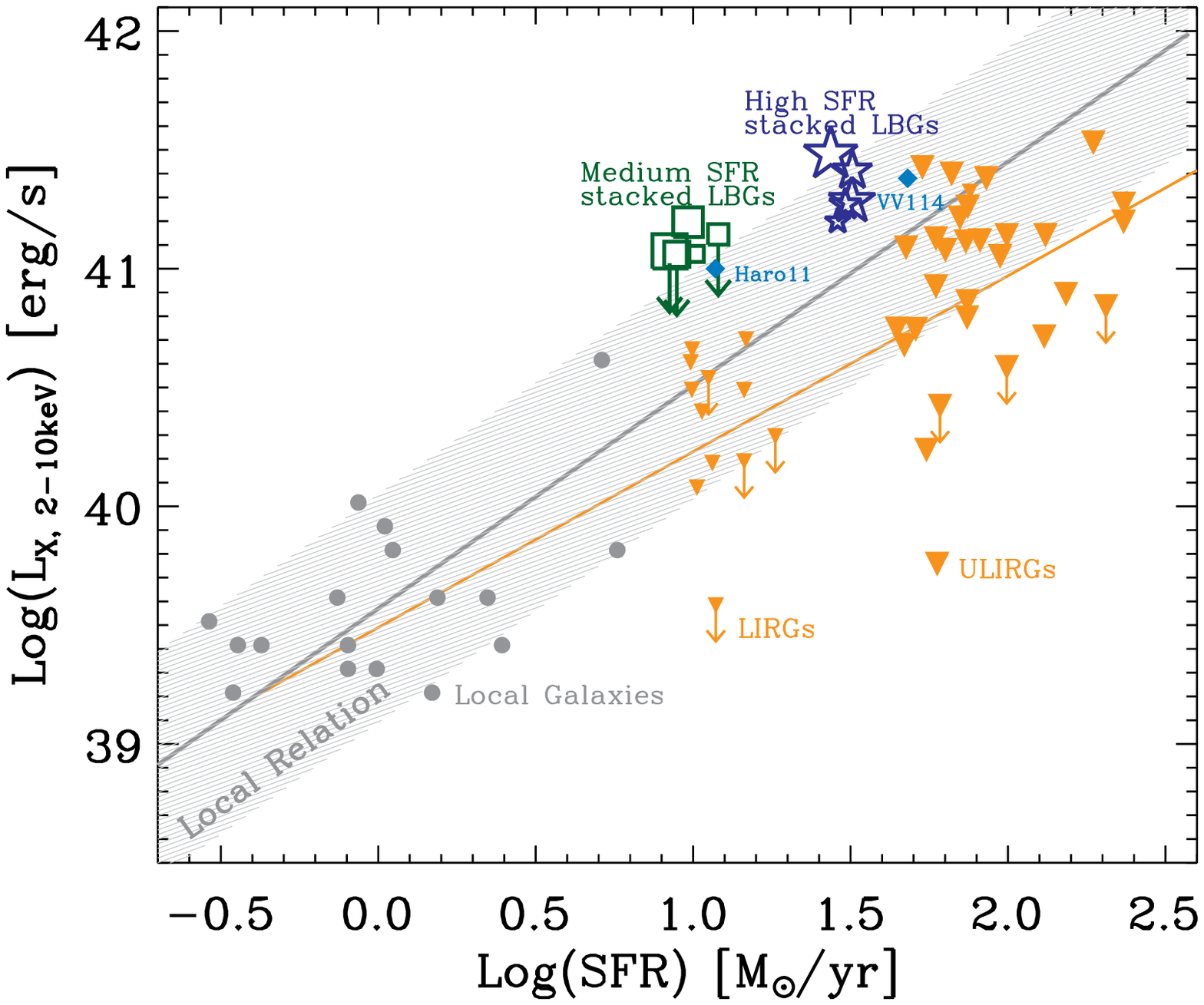}
    \end{center}
  \caption{We compare the 2--10~keV (rest-frame) luminosity versus SFR for different galaxy populations to show that UV-selected galaxies appear to deviate towards higher L$_{\rm X}$.  Local galaxies (gray points, from \citealt{Colbert04}) and LIRGs/ULIRGs (orange triangles from \citealt{iwasawa} and \citealt{Lehmer2010}). Stacked medium SFR LBGs (open green squares), stacked high SFR LBGs (open blue stars), two local analogs (cyan diamonds, Haro11; \citealt{grimes07} and VV114; \citealt{grimes06}) all have higher X-ray luminosities per SFR compared to other galaxy populations. The LBG symbol sizes correspond to the redshifts, smallest ($z=1.5$) to largest ($z=3.8$). Solid gray line (shaded region) shows correlation (scatter) based on low SFR galaxies and solid orange line shows fit to data, including IR-selected galaxies, from \citealt{Lehmer2010}. }
   \label{fig:lx_sfr}
   \end{figure}
   
\begin{figure*}
\begin{center}
  \includegraphics[width=3.5in]{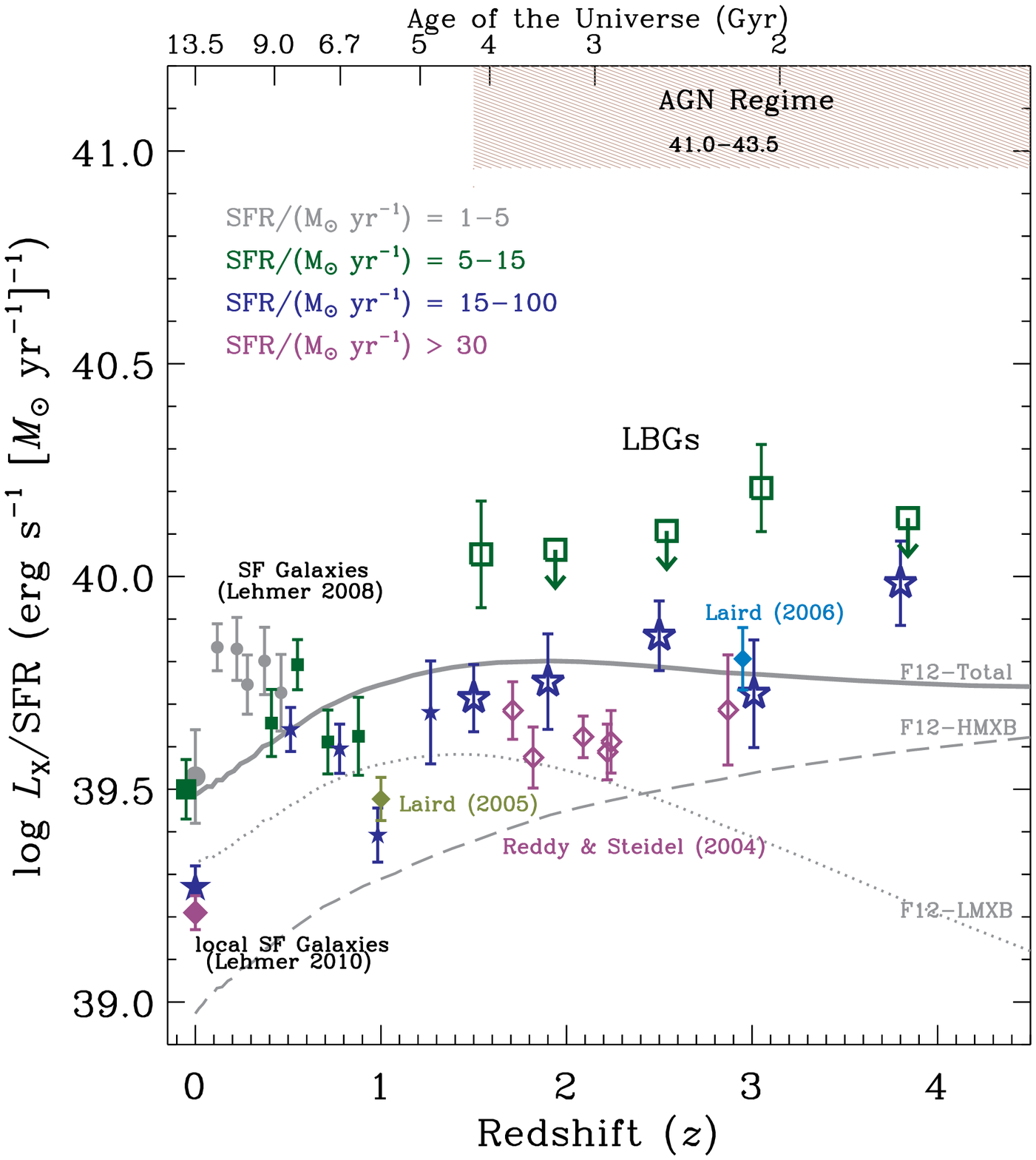}
  \includegraphics[width=3.5in]{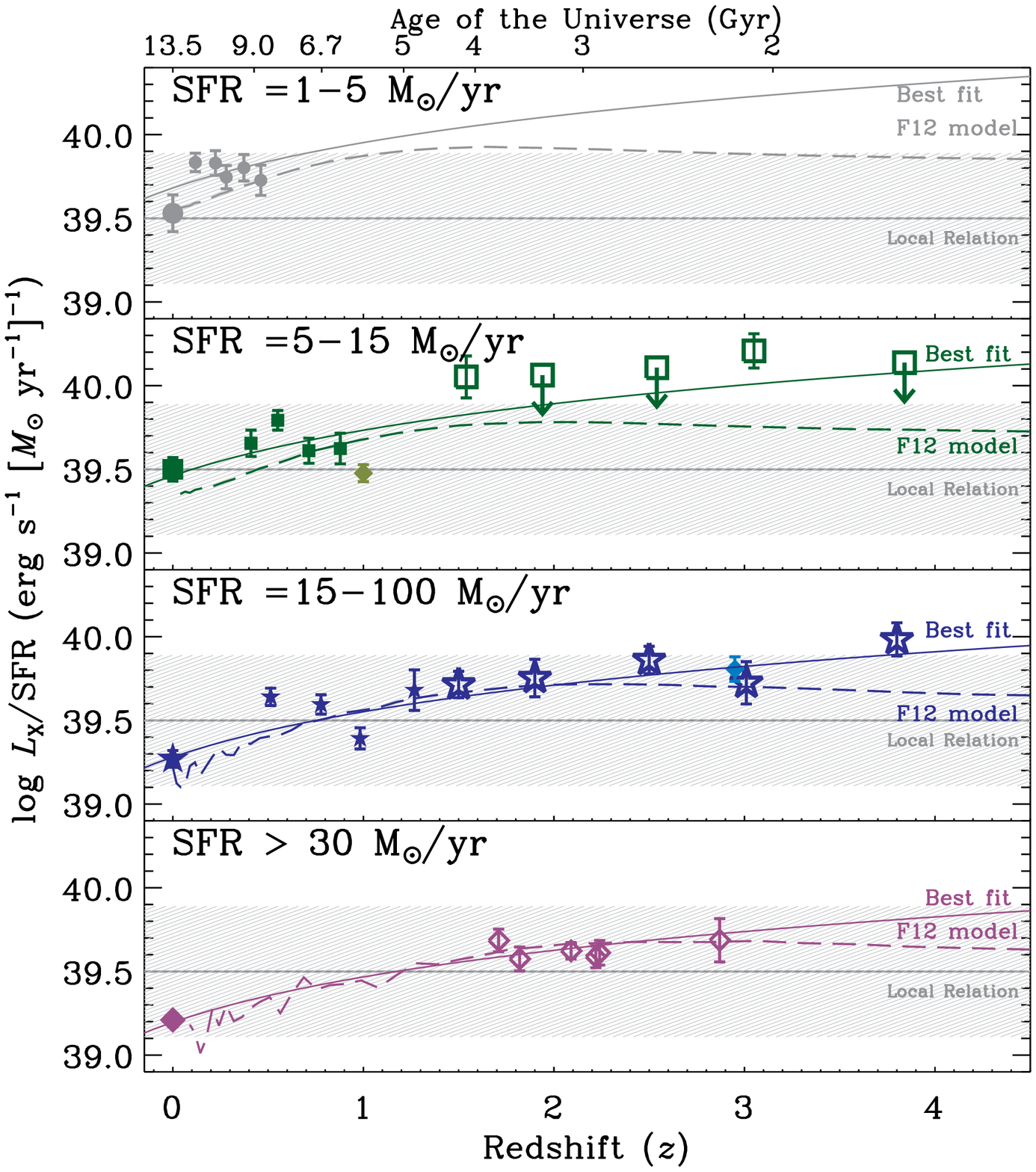}
    \end{center}
  \caption{We show the evolution of rest-frame hard (2--10 keV) X-ray luminosity per SFR between $0<z<4$ (the last 12 Gyr of cosmic history). The {\em left} panel shows all the data, while the {\em right} panel separates the results into different SFR ranges to show redshift trends. Symbols mark bins of galaxies with different SFR ranges: gray circles have low SFRs, green squares show medium SFRs, blue stars show high SFR galaxies, and purple diamonds mark the highest SFR galaxies (SFR/\msunyrs$>30$). Following the same scheme, larger filled symbols show local ($z\sim0$; \citealt{Lehmer2010}) galaxies, and smaller filled symbols are star-forming galaxies between $0<z<1.4$ (\citetalias{L08}). Stacking results from other star-forming galaxies are shown: $z=1$ Balmer break galaxies \citep[filled green diamond;][]{Laird05},  $z=$1.5--3 BX/BM galaxies \citep[open purple diamonds;][]{Reddy04}, and $z=3$ LBGs \citep[filled cyan diamond;][]{Laird06}.  {\em Left:} The red hatched region shows the regime, $\log$ L$_{\rm X}=41.0-43.5$ \ergs, inhabited by individually detected X-ray sources, assumed to be AGN. The gray curves show X-ray binary synthesis models from \citetalias{F12} for total (LMXB$+$HMXBs; solid), HMXBs (dashed) and LMXBs (dotted) for galaxies with SFRs $>$1 \msunyrs. {\em Right:} The gray line and shaded region (shown in all panels) represent the local X-ray/SFR relation and its scatter (derived by \citetalias{L08}). The best fit model parameterization (solid curves) show weak redshift evolution, as described by Equation \ref{eqn:fit}. X-ray binary synthesis models from \citetalias{F12} (dashed curves) are not fits to the data and show excellent agreement with the data and our best-fit. }
   \label{fig:lxsfrVz}
   \end{figure*}
   
   \begin{figure}
\begin{center}
\hspace{-0.4in}
  \includegraphics[width=3.7in]{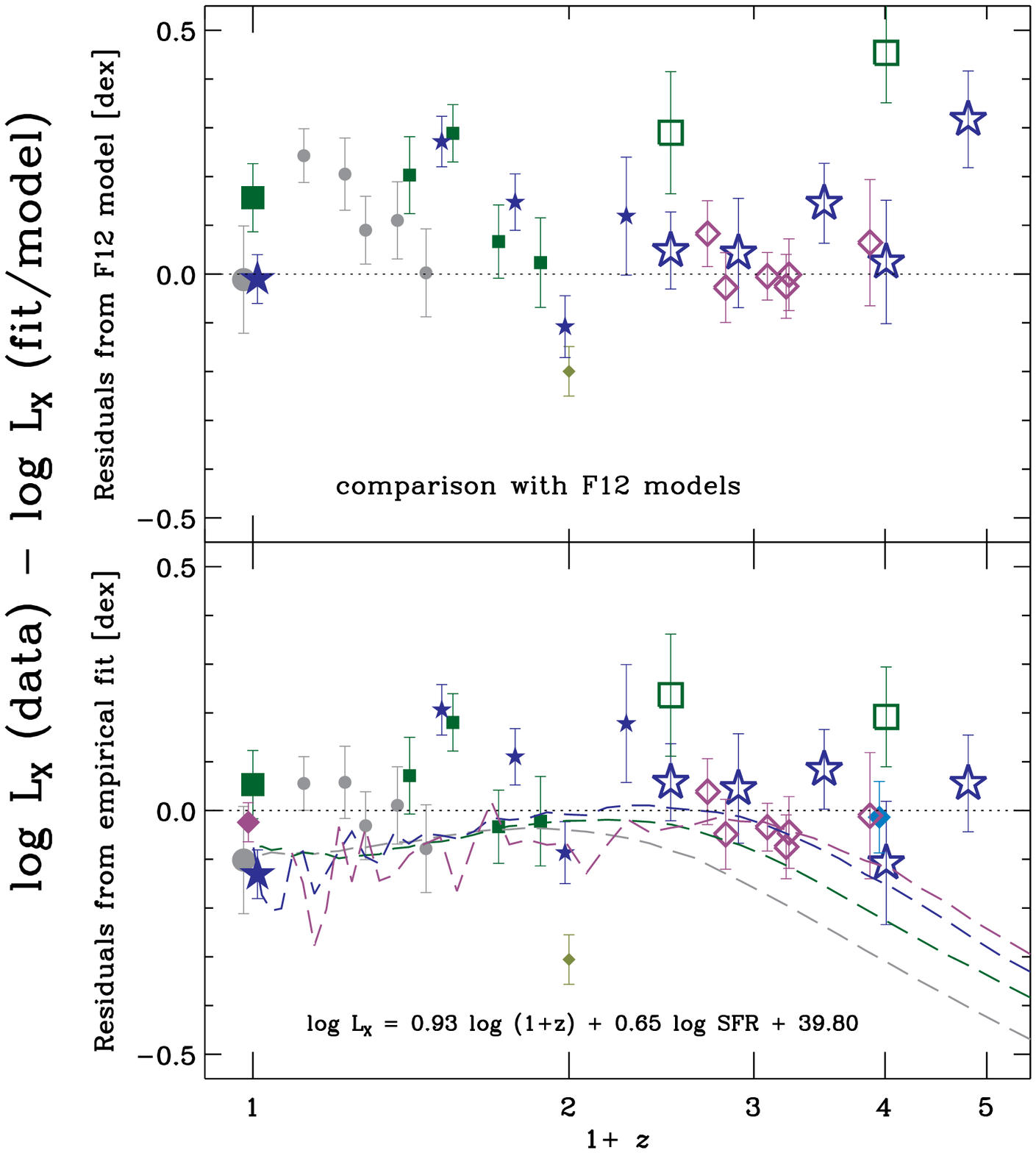}
    \end{center}
    \vspace{-0.2in}
  \caption{We compare residuals between data and X-ray binary synthesis models from \citetalias{F12} (top panel) and empirical fits (bottom panel) versus redshift. The colors and symbols are as described for Figure \ref{fig:lxsfrVz}. {\em Top panel}: \citetalias{F12} models agree excellently with the data at all redshifts but best for the low SFR and high SFR samples at $z=0$ and the SFR $>30$ \msunyr at $z>2.5$. {\em Bottom panel}: We display the residuals of the empirical fit ($\log$ L$_{\rm X}$ is fit by the equation stated here) to the data (symbols) and \citetalias{F12} models (dashed curves; with the colors corresponding to the SFR bins as in the previous Figure: gray, green, blue and purple represent SFR/\msunyr $=1-5$, $5-15$, $15-100$ and $>$30, respectively). 
}
   \label{fig:lxsfr_model}
   \end{figure}
\subsection{X-rays and star-formation}\label{sec:Xray-sf}
Concentrating on the X-ray properties related to star-formation in LBGs, we split our five $z\lesssim4$ LBG samples into two separate SFR bins for stacking: ${\rm SFR}/($\msunyr$)= 5-15$ (medium; see Figure \ref{fig:SFRhist}) and ${\rm SFR}/($\msunyr$)=15-100$ (high). We do not include the low SFR galaxies (SFR $=1-5$ \msunyr) in our analysis because the LBG samples suffer from incompleteness here (see Figure \ref{fig:SFRhist}) and there are no stacked detections, with upper limits providing only weak constraints. Stacking of the high SFR LBGs all yield detections (with 2--10 keV X-ray luminosities ranging from 1.6--3.0$\times 10^{41}$\ergs), while only the $z\approx1.5$ and 3 samples are detected for the medium SFR bins. The results from this analysis are summarized in Table \ref{tab:stackSF}. Our least significant detection (at 2.6-$\sigma$) is the $z=1.5$ sample, whose stacked 0.5--2 keV image is shown on the left side of Figure \ref{fig:imstack}. Within each redshift and SFR bin, the mean and median values of the SFRs (given in columns 6 and 7 of Table \ref{tab:mass}) agree fairly well, suggesting that the X-ray stacking results are representative for the whole sample and not skewed by galaxies in a particular SFR regime.

In Figure \ref{fig:lx_sfr}, we show the X-ray/SFR relation (L$_{\rm X}$ vs. SFR) for the stacked LBGs (open blue stars; sizes relate to redshift, smallest to largest range from $z=1.5-$3.8) as compared to other samples of star-forming galaxies: local galaxies from \citet{Colbert04}, luminous infrared and ultraluminous infrared galaxies (LIRGs/ULIRGs; orange triangles) from \citet{Lehmer2010} and \citet{iwasawa}. 
VV114 and Haro11 are two local LBG analogs \citep{grimes06,grimes07} and shown as labelled cyan diamonds in Figure \ref{fig:lx_sfr}. The stacked LBGs appear to have a similar X-ray/SFR ratio to the low SFR local galaxies (gray points). However, all of the UV-selected samples (the local analogs, VV114 and Haro11, and the stacked $z=1.5-4$ LBGs) appear to have slightly higher L$_{\rm X}$ per SFR. 
The gray line (and shaded region) follow the L$_{\rm X}$ versus SFR relation (and 1-$\sigma$ scatter) derived in \citet{Lehmer2010} for the gray points, (hereafter, the local relation). The high SFR end of this relation is difficult to study since few local galaxies have such high SFRs. However, \citet{Lehmer2010} find that the infrared-selected galaxies (LIRGs/ULIRGs) deviate from the local relation, with lower L$_{\rm X}$ per SFR (\ie they are better fit by the orange solid line). They argue that in typical galaxies, both HMXBs and LMXBs provide substantial contributions to the total X-ray luminosity, while high specific SFR galaxies have X-ray emission dominated by HMXBs alone, thereby causing a lower L$_{\rm X}$/SFR (see equation \ref{eqn:lx}). Studying the X-ray/IR correlation in  $0<z<2$ galaxies, \cite{Symeonidis11} also find that the correlation deviates from the local correlation at high L$_{\rm IR}$, presumably due to relatively high specific SFRs. Other local ($z<0.2$) samples of UV-selected galaxies, with high specific SFRs, can offer important insight at these high SFRs and further investigation of this relation at the high SFR end is ongoing \citep{me-chandralba}.  

While UV-selected galaxies (locally and at $z>1.5$) appear similarly shifted in the X-ray/SFR relation compared to other galaxy populations, it has not been firmly established whether the local X-ray/SFR relation can be applied to high redshift galaxies. \citet{Cowie11} claim that the ratio of X-ray luminosity to UV luminosity does not evolve from the local relation, if you assume a constant dust attenuation factor (with values between 3--5) for stacked $z=$1--6 LBGs. In Figure \ref{fig:lx_sfr}, the stacked LBGs (open green squares and blue stars) show subtle variation with redshift (symbol size ranges from $z=1.5$, smallest, to $z=4$, largest). However, extending our analysis to $z<1.5$ star-forming galaxies and taking dust attenuation corrections into account, we further examine the X-ray/SFR relation for traces of evolution.

In Figure \ref{fig:lxsfrVz}, we show the mean L$_{\rm X}$/SFR ratio of our sample and further include results from \citet[][filled green diamond]{Laird05}  for SFR$=$5--15 \msunyrs, \citet[][filled cyan diamond]{Laird06} for SFR$=$15--100 \msunyrs, and \citet[][open purple diamonds]{Reddy} for SFR$>30$ \msunyrs. For comparison with lower redshift samples, we include the local star-forming galaxies from \cite{Lehmer2010} and the $0<z<1.4$ star-forming galaxies from \citetalias{L08}. We note that the high SFR point at $z=0$ from \cite{Lehmer2010} is dominated by IR-selected galaxies (as discussed previously, these appear to have lower L$_{\rm X}$ per SFR). 

Combining the information from the data shown in Figure \ref{fig:lxsfrVz}, including our LBGs sampled by redshift and SFR bins (given in Table \ref{tab:stackSF}) and other published data, we parameterize the X-ray luminosity in terms of redshift and SFR. Using error-weighted least-squares fitting to the stacked detections, we derive the following best fit to the available data: 
\begin{eqnarray}
\log {\rm L}_{\rm X}& =& A~\log (1+z)+ B\log{\rm SFR}+C \label{eqn:fit}\\
A & = &0.93\pm{0.07}\notag \\
B & = & 0.65\pm{0.03}\notag \\
C & = & 39.80\pm{0.03} \notag
\end{eqnarray}
The trend for increasing L$_{\rm X}$/SFR with decreasing SFR is consistent with what is observed for local galaxies. For example, \citet{Lehmer2010} find L$_{\rm X}$ varies with SFR in a similar way (within uncertainties, referring to coefficient B in their Table 4) for their local sample of LIRGs. 

In addition, we find evidence of weak redshift evolution (shown as solid curve in right panel of Figure \ref{fig:lxsfrVz}, where SFR in the equation is set to the mean SFR of the displayed sample). Our result (Equation \ref{eqn:fit}) is robust even when we restrict our analysis to the most complete samples (high SFR LBGs), where selection effects (\ie flux limits) are minimal. Therefore, we argue that this evolution is driven by that of physical properties (\eg metallicity, star formation history) within the galaxies. 

   \begin{figure}
\begin{center}
  \includegraphics[width=3.5in]{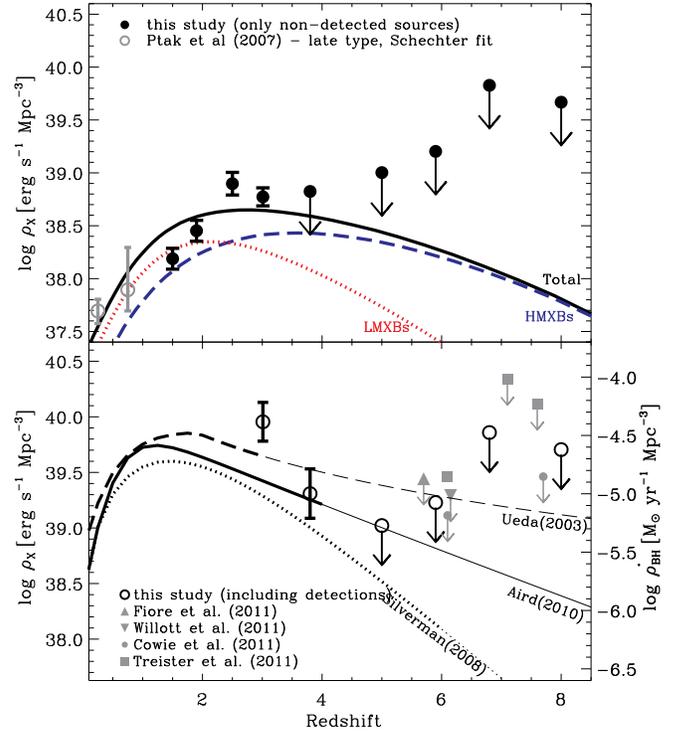}
    \end{center}
    \vspace{-0.3in}
  \caption{{\em Top:} The observed evolution of the normal-galaxy XLD, $\rho_{\rm X}$, from our stacked LBGs (solid black points) and $z<1.4$ late-type galaxy X-ray luminosity function (open gray circles; \citealt{Ptak07}) shows excellent agreement with XRB models (solid black, dashed blue and red dotted lines for total, HMXBs and LMXBs, respectively; for SFR$>$1\msunyr galaxies). {\em Bottom:} Open circles show total $\rho_{\rm X}$ (including individually detected sources) from stacking, which are primarily dominated by X-ray emission from AGN. Results from other stacking analyses are shown in gray symbols, as described in the legend. The curves show X-ray luminosity function models, as labelled, where thinner lines show extrapolated values. The axis on the right-side provides black hole mass accretion rates (see text for details). Since no individually detected LBGs were present in the $z\gtrsim 6$ sample, the solid (top panel) and open (bottom panel) black circles are identical at these redshifts.}
   \label{fig:BHgrowth}
   \end{figure}
\citetalias{F12} perform a large scale population synthesis study, using the {\tt StarTrack} binary population synthesis code \citep{Belczynski2002, startrack}, that models XRB populations from the first galaxies of the Universe until today. They use as input to their modeling the Millennium II Cosmological Simulation \citep{MS-II} and the updated semi-analytic galaxy catalog by \citet{Guo2011} to self-consistently account for the star formation history and metallicity evolution of the universe. Their models, which are constrained by the observed X-ray properties of local galaxies, give predictions about the global scaling of emission from X-ray binary populations with properties such as SFR and stellar mass, and the evolution of these relations with redshift. 

The results presented in \citetalias{F12} correspond to a general population of galaxies, including a mixture of actively star-forming and passive early type galaxies. In this paper, we consider the maximum likelihood model, as reported in \citetalias{F12}, which we adapt to the galaxy sample properties of our survey. More specifically, while \citetalias{F12} used all the galaxies in the Millennium II simulation box in order to derive the star-formation history and metallicity evolution, here we only took into account the evolution of galaxies with SFR in four bins (1--5, 5--15, 15--100, $>$30 \msunyrs). We assume that LBGs are representative of the galaxy population over these redshifts ($z=$1--5) and therefore any metallicity and star formation history evolution in the LBGs will also be described by galaxies in the Millenium II simulation (and therefore, in the \citetalias{F12} models). These adapted models can be directly compared to our stacking results (see Figures \ref{fig:lxsfrVz}--\ref{fig:BHgrowth}). 

Noting that these X-ray binary synthesis models are not fits to our data, there is remarkable agreement of these models with the data and empirical fits. The top panel of Figure \ref{fig:lxsfr_model} shows the residuals between the \citetalias{F12} models and data versus redshift. Residuals between the empirical fit and the data are shown in the bottom panel of Figure \ref{fig:lxsfr_model}. The dashed curves show the differences between this best-fit relationship and the \citetalias{F12} models, with the colors corresponding to SFR bins as in the previous Figure: gray, green, blue and purple represent SFR/(\msunyr) $=$1--5, 5--15, 15--100 and $>$30, respectively. 

We use the \citetalias{F12} models to interpret the evolution in the X-ray/SFR relation. The evolution of 2--10~keV emission per SFR from XRBs involves several competing effects: star formation history, the evolution of XRB populations, and metallicity evolution. The contributions of HMXBs and LMXBs shift with the global star formation history of the Universe; HMXBs trace the young stellar populations and are expected to scale with SFR, while LMXBs track older stellar populations which dominate once the young stars are extinct. This effect can be seen in the shape of the LMXB curve (dotted gray) in Figure \ref{fig:lxsfrVz} [{\em left}], which includes the global star formation history. Metallicity affects the number of HMXBs and their X-ray luminosities; high mass stars, which eventually evolve into black hole HMXBs (more X-ray luminous than neutron star HMXBs) are easier to produce in high metallicity environments because of weaker stellar winds \citepalias{F12}. At higher redshifts, metallicities are lower and the contribution of HMXBs to the X-ray luminosity is higher for a given SFR, apparent from the gray dashed line (HMXB) in Figure \ref{fig:lxsfrVz} [{\em left}]. The combination of these effects on the X-ray luminosity per SFR is shown as the solid gray line in left panel of Figure \ref{fig:lxsfrVz}. With additional information (\eg accurate stellar masses, metallicities and star formation histories), we could further explore how well the X-ray binary synthesis models describe these separate effects in observed galaxies. However, in the absence of such data, we interpret the evolution of X-ray luminosity per SFR to be driven by the metallicity evolution of HMXBs. Given their high sSFRs (10$^{-8.7}$--$10^{-8.0}$, consistent with containing young stellar populations), HMXBs are likely to dominate the X-ray emission in LBGs; metallicity evolution in HMXBs causes an increase in the L$_{\rm X}$/SFR with redshift. 

\subsection{Evolution of X-ray luminosity density as related to X-ray binaries}\label{sec:Xray-XLD}
To determine the total X-ray contribution of X-ray binaries in the Universe, we stack the full samples of LBGs (including all SFRs, but excluding X-ray detected sources). The top panel of Figure \ref{fig:BHgrowth} provides our measurements of the 2--10 keV X-ray luminosity density (XLD), $\rho_{\rm X}$, due to star-formation from LBGs. $\rho_{\rm X}$ was calculated in the following way:
\begin{eqnarray}
\rho_{\rm x} & = & n_{\rm LBG}\langle \rm L_{\rm X} \rangle \\
 & = &\rm \langle L_{X} \rangle \int_{\rm L_{min}}^{\infty}\phi(L_{UV}) dL_{UV}
\end{eqnarray}
by multiplying the average (stacked) rest-frame 2--10~keV X-ray luminosity, $\langle L_{\rm X} \rangle$, by the LBG number density (by integrating the LBG luminosity function, $\phi(\rm L_{UV})$, down to the faintest observed UV magnitude limit for the sample; refer to papers listed in Table \ref{tab:LBGs} for luminosity function fitting parameters). We also include $\rho_{\rm X}$ values determined from the X-ray luminosity function, fit by a Schechter function, of late-type, low redshift ($z<1.4$), galaxies \citep[][converted from 0.5--2~keV band into 2--10~keV using a power law SED with $\Gamma=2$]{Ptak07}, shown as gray open circles. We compare the measured values to those predicted by \citetalias{F12} based on total X-ray binary contribution (black) and separated into contributions from HMXBs (dashed blue) and LMXBs (dotted red). In general, there is good agreement of the total X-ray binary curve with the data; the models include all galaxies with SFRs $>1$ \msunyrs, while the minimum SFR varies slightly between redshift samples in the data (see Figure \ref{fig:SFRhist}). For example, at $z\approx1.5$, the measured value appears slightly lower than the expected $\rho_{\rm X}$ from HMXBs $+$ LMXBs, which may be caused by incompleteness in the $z\approx1.5$ LBG sample. Another explanation is that $z=1.5$ LBGs may not be representative of all high SFR ($>1$ \msunyr galaxies; \ie the LBG selection would miss dusty star-forming galaxies like LIRGs or low sSFR galaxies dominated by older stellar populations). The $z\approx 2.5$ measured value appears slightly higher than the model, and may include contamination from low luminosity AGN. 

\begin{deluxetable*}{lccccccc}
\tabletypesize{\scriptsize}
\setlength{\tabcolsep}{0.02in}
\tablecolumns{8} 
\tablewidth{6.5in} 
\tablecaption{Properties of Stacked $1.5<z<8$ LBGs Related to X-ray Luminosity Density} 
\tablehead{
\colhead{$z$} & \colhead{\# of sources} & \colhead{t$_{tot}$} &  \colhead{Net counts} &  \colhead{S/N} & \colhead{$\log$ L$_{\rm X}$\tablenotemark{a}} & \colhead{$\log$ L$_{\rm UV}$} &    \colhead{$\log \rho_{\rm X}$\tablenotemark{b}}  \\
 &  &  \colhead{(Ms)} & & &\colhead{(erg s$^{-1}$)} & \colhead{(erg s$^{-1}$)}  & \colhead{(erg s$^{-1}$ Mpc$^{-3}$)}
 }
\startdata
  \cutinhead{Total X-ray from X-ray binaries (No SFR cut; excluding individually detected X-ray sources\tablenotemark{$\dagger$})}
    1.5   &     30   &  120.0   &      56.26 $\pm$ 14.36   &        4.2   &     41.06   &  44.12 &       38.19$\pm$ 0.10     \\
  1.9   &     44   &  176.0   &      69.16 $\pm$ 17.69   &        4.3   &     41.20   &  44.17 &       38.45$\pm$ 0.10     \\
  2.5   &    222   &  888.0   &     117.68 $\pm$ 32.89   &        3.3   &     41.01   &  44.15 &       38.90$\pm$ 0.11     \\
  3.0   &    311   & 1244.0   &     226.38 $\pm$ 49.08   &        5.4   &     41.21   &  44.50 &       38.77$\pm$ 0.09     \\
  3.8   &   1381   & 5524.0   & $<$ 222.48               &        1.6   & $<$ 40.82   &  44.12 &   $<$ 38.82               \\
  5.0   &    701   & 2804.0   & $<$ 158.51               &        0.9   & $<$ 41.18   &  44.18 &   $<$ 39.00               \\
\cutinhead{Total X-ray from all sources (including individually detected X-ray sources\tablenotemark{*})}
  3.0   &    319   & 1276.0   &    3513.55$\pm$ 1736.9   &       82.1   &     42.39   &  44.52 &       39.96$\pm$      0.17     \\
  3.8   &   1388   & 5552.0   &     667.99$\pm$  445.7   &        7.5   &     41.31   &  44.12 &       39.31$\pm$      0.22     \\
  5.0   &    702   & 2808.0   & $<$ 158.62               &        2.0   & $<$ 41.20   &  44.18 &   $<$ 39.02                    \\
  5.9   &    220   &  880.0   & $<$  88.56               &        1.0   & $<$ 41.62   &  44.10 &   $<$ 39.23                    \\
  6.8   &     61   &  244.0   & $<$  47.07               &       -1.0   & $<$ 41.98   &  44.18 &   $<$ 39.86                    \\
  8.0   &     44   &  176.0   & $<$  40.03               &       -1.8   & $<$ 42.20   &  44.12 &   $<$ 39.71                    \\
  \enddata
  \tablenotetext{a}{Total luminosity in 2--10 keV (rest-frame) band.}
  \tablenotetext{b}{Total luminosity density in 2--10 keV (rest-frame) band.}
  \tablenotetext{$\dagger$}{Since $z=5.9, 6.8$ and 8.0 have no individually detected X-ray sources, the information for these redshifts is the same as given below. }
  \tablenotetext{*}{We exclude $z=1.5, 1.9$ and 2.5 samples from this analysis because the volume probed by those samples is small (see areal coverage map in Figure \ref{fig:cdfs-field}) and since luminous X-ray sources are rare, any analysis regarding the total X-ray luminosity probed by LBGs in this redshift range is too incomplete to yield meaningful results.}
\enddata
\label{tab:stackBH} 
\end{deluxetable*}
\subsection{$z\approx3-8$ constraints on the evolution of supermassive black holes}\label{sec:Xray-BH}
At $z\gtrsim 5$, we are unable to obtain detections by stacking only sources that are individually X-ray undetected. However, by including individually detected X-ray sources, we can provide constraints on black hole growth in these galaxies. Table \ref{tab:stackBH} summarizes the results from our $z\gtrsim 5$ stacking. The top section of Table \ref{tab:stackBH} excludes individually detected X-ray sources, while the bottom section lists the results from stacking all the sources. We note that including X-ray detected sources in our stacking analysis at $z\gtrsim 5$ only affects the measurement at $z\approx5$ (slightly, because of one X-ray detected source). Therefore, while the results at $z\gtrsim 6$ remain unchanged, we shift our focus from studying accretion in X-ray binaries to that in supermassive black holes (SMBHs). 

In the bottom panel of Figure \ref{fig:BHgrowth}, we show our measurements of the 2--10~keV XLD, $\rho_{\rm X}$, from LBGs. $\rho_{\rm X}$ was calculated by multiplying the average (stacked) X-ray luminosity, $\langle L_{\rm X} \rangle$, by the number density (as described in Section \ref{sec:Xray-sf} for the top panel of this figure). Here, the X-ray luminosities have been calculated assuming $\Gamma=1.8$, which is appropriate for AGN (whereas, all of our previous calculations have been using $\Gamma=1.7$, see Section \ref{sec:stack}). This observed quantity, $\rho_{\rm X}$, can be used to estimate a black hole mass accretion rate density ($\rho_{\rm BH}$, shown in log units on axis at right side) for the Universe by assuming a black hole radiation efficiency ($\epsilon$) and bolometric correction ($C_{bol}$) as:
\begin{equation}
\dot \rho_{\rm BH} = \frac{(1-\epsilon)}{\epsilon {\rm c}^2} C_{bol} \rho_{\rm X}~{\rm M_\odot yr^{-1} Mpc^{-3}}
\end{equation}
where c is the speed of light, $\rho_{\rm X}$ is given on the axis on left, and we adopt the values $\epsilon=0.1$ and $C_{bol}=$25 following \citet{Treister11}. In this figure, we show the total (including the single individually detected source, listed in Table \ref{tab:det}) stacked X-ray luminosity per Mpc$^{-3}$ as open black circles. To compare with other studies, we use the 0.5-2 keV count rates\footnote{where 3$\sigma$ upper limits are given, we converted to 2.5$\sigma$ upper limits to the count rates for consistency with our analysis.} from other high redshift stacking studies to convert consistently to X-ray luminosities as described in Section \ref{sec:stack}; these results are shown as the gray upward triangle \citep{Fiore11}, downward triangle \citep{Willott11}, small circle \citep{Cowie11} and square symbols \citep{Treister11}. 

We note these other studies have more sources in their stack since they limit sources with off-axis angles greater than 9\arcmin~versus our adopted  7\arcmin~off-axis limit; however we have checked that using the 9\arcmin~limit does not change the results of our stacking analysis by much ($<10\%$). In agreement with \citet{Fiore11}, \citet{Cowie11}, and \citet{Willott11}, we do not obtain a detection for the stacked $z\approx 5.9$ sample (see {\it right} image in Figure \ref{fig:imstack}), which was previously reported in \citet{Treister11}. Rather our upper limits for the $z\approx 6$ and 7 data are very similar to those of \citet{Fiore11}. \citet{Willott11} uses an optimized aperture and weighting in order to push their upper limits slightly lower than the others\footnote{Since \citet{Fiore11} and \citet{Willott11} do not include CANDELS LBGs, their upper limits are not as constrained as ours. But when we exclude the CANDELS data, our upper limits are the same as theirs.}. At $z\approx 6$, our upper limit is lower than the 5$\sigma$ detection claimed by \citet[][though this detection has been called into question in other papers; see \citealt{Willott11,Cowie11}]{Treister11}. The solid, dashed, and dotted lines show the 2--10 keV XLD evolution model from \citet{Aird2010}, \citet{Ueda03}, and \citet{Silverman08}, extrapolated beyond the measured redshifts (at $z=4$, 3.0, and 6.0, respectively; displayed in thinner lines) assuming the same form and derived parameters. Our observations are most consistent with the XLD evolution of AGN with L$_{\rm X}>10^{42}$ \ergs~ as predicted by \cite{Aird2010} and less consistent (based on the upper limits at $z\approx 5-6$) with \citet{Ueda03}. Yet, we are not able to definitively rule out the \cite{Ueda03} models since extrapolation to higher redshifts ($z>3$) may not be accurate, and at $z=3$ the observations are consistent with their model. Additionally, calculating $\rho_{\rm X}$ from AGN in LBGs alone may underestimate the XLD from the entire AGN population, present also in populations of galaxies not selected by the LBG selection (\eg submillimeter galaxies). We may be underestimating the contributions from Compton-thick AGN: \cite{Gilli2011} detect a Compton-thick AGN at $z=4.76$, suggesting that Compton-thick AGN may be important (though difficult to detect and study) for understanding black hole evolution in the early universe. 

\section{Summary and Future Work}\label{sec:end}
Using the deepest X-ray survey to date, the CDF-S 4 Ms data, and the most comprehensive LBG catalogs, we have stacked $1.5<z<8$ LBGs. We have split our analysis into two separate redshift ranges: $z\lesssim4$ to study the relation between SFR and X-rays produced by X-ray binaries, and $z\gtrsim 5$ to place constraints on the black hole accretion history of the universe. Our main results are as follows:
\begin{enumerate}
\item{We expand the list of X-ray detections (found in the \citetalias{X11} catalog) associated with LBGs (see Table \ref{tab:det}), based on probabilistic matching between X-ray and optical/infrared catalogs. Within 7\arcmin~of the \Chandra pointing, we match 20 LBGs with individually detected X-ray sources, assumed to be AGN based on their X-ray luminosities ($>10^{42}$~\ergs)}.
\item{While the local X-ray/SFR relation does appear to apply to the high redshift LBGs, it also has significant scatter. We find that the population of UV-selected galaxies (including local LBG analogs, also selected by using the rest-frame UV) preferentially lie at the high end of the X-ray/SFR correlation scatter (see Figure \ref{fig:lx_sfr}). The high SFR regime of this relation has not been well-studied for different samples of galaxies, but these observations deviate from what is observed for high SFR, IR-selected LIRGs and ULIRGs \citep{Lehmer2010, Symeonidis11}. }
\item{We discover mild evolution of L$_{\rm X}$ per SFR between $0<z<4$ (see Figure \ref{fig:lxsfrVz}). We stacked the LBGs in two different SFR ranges: 5--15 \msunyr  (medium) and 15--100 \msunyr (high). We obtain $>2.5\sigma$ detections in the stacked high SFR bins, and for two bins ($z=1.5$ and 3) in the stacked medium SFR samples. We compare our results for the high SFR galaxies with XRB population synthesis models \citepalias{F12} and find excellent agreement (see Figure \ref{fig:lxsfr_model}). We find that the rest-frame 2--10~keV X-ray luminosity relates to SFR and redshift in the following way: $\log$ L$_{\rm X} = 0.93\log(1+z)+0.65\log{\rm SFR}+39.8$. }
\item{We calculate the total (including all SFRs, but excluding individually detected X-ray sources) XLD for different redshift samples to compare with XRB population synthesis models \citepalias[][see top panel of Figure \ref{fig:BHgrowth}]{F12}. We find that HMXBs and total (HMXBs +LMXBs) are both consistent with the data. }
\item{Stacking $z\gtrsim 5$ LBGs did not provide any detections (see bottom panel of Figure \ref{fig:BHgrowth}). However, we use the upper limits to derive the 2--10 keV XLD and we compare with results from other studies and models \citep{Aird2010, Silverman08, Ueda03}. Our results are similar to \citet{Fiore11}, \citet{Cowie11}, and \citet{Willott11}: we do not detect LBGs at $z=6$ \citep[contrary to][]{Treister11}. Our observations are consistent with the models from \citet{Aird2010} and \citet{Silverman08}, extrapolated to $z=8$; our upper limit at $z\sim5$ lies below the extrapolated ($z>3$) \citet{Ueda03} model. However, we note that our comparisons with XLD evolution models are subject to the following caveats: extrapolation of the XLDs beyond the redshifts for which they were measured may not be accurate; studying AGN in LBGs may not account for the complete AGN population at these redshifts (\eg AGN in submillimeter galaxies); we may be underestimating the contribution from Compton-thick AGN.     }
\end{enumerate}

Future missions like the James Webb Space Telescope (JWST) aim to discover the first galaxies, when the Universe was only a few 100 million years old, searching for very high redshift ($z>10$) LBGs. Unfortunately, studying such distant LBGs in the X-rays would be impossible with current X-ray telescopes, and X-ray instruments capable of such ambitious ventures exist only the imagination, as yet. However, the future does offer other opportunities for studying X-ray emission from distant LBGs. 

\cite{Fiore11rev} describe the future prospects for studying AGN in LBGs using future X-ray facilities like Athena and Super-Chandra. As for studying XRB populations in these galaxies, deeper X-ray observations would add stacked detections (where we now have upper limits) and better constrain the results presented in this paper. We estimate, using our Equation \ref{eqn:fit}, that a 5~Ms \Chandra exposure (\ie adding 1~Ms to the current CDF-S observations) would provide stacked detections ($>3\sigma$) for all the $z\approx 2.5-3.8$ medium SFR sample, and 6--8~Ms \Chandra exposure would provide a stacked detection for the 100 highest SFR LBGs at $z\approx5$. Deep IR observations from JWST and {\it Hershel} would provide additional rest-frame optical and IR photometry to provide better redshift determinations for the current LBGs at $z=5-8$, improving the purity of the LBG selection and offering more accurate measurements of stellar masses, dust attenuations, star formation histories, and SFRs for individual LBGs. X-ray stacking by these properties would provide insight about how the evolution of fundamental properties with redshift relate to the evolution of X-ray emission in these galaxies. 

\begin{acknowledgments}
We are grateful to the Chandra Director's office for commissioning the 4 Ms observation of the CDF-S. We thank Ezequiel Treister for helpful discussions and the anonymous referee for his/her suggestions. This research was supported by Chandra Cycle 12 program \#12620841 (P.I.: Basu-Zych) and NASA ADP Proposal 09-ADP09-0071 (P.I.: Hornschemeier). A.R.B. gratefully acknowledges the appointment to the NASA Postdoctoral Program at the Goddard Space Flight Center, administered by Oak Ridge Associated Universities through a contract with NASA, and NASA's {\it Swift} Observatory for salary support. We gratefully acknowledge financial support from the Einstein Fellowship Program (B.D.L.), the Youth 1000 Plan program and the USTC startup funding (Y.Q.X.). P.O. acknowledges support provided by NASA through Hubble Fellowship grant HF-51278.01. T.F. acknowledges support from the CfA and the ITC prize fellowship programs. W.N.B., B.L., and Y.Q.X. thank CXC grant SP1-12007A and NASA ADP grant NNX10AC99G. Herschel is an ESA space observatory with science instruments provided by European-led Principal Investigator consortia and with important participation from NASA. The GOODS-H data was accessed through the HeDaM database ({\ttfamily http://hedam.oamp.fr}) operated by CeSAM and hosted by the Laboratoire d'Astrophysique de Marseille.
\end{acknowledgments}
\bibliographystyle{apj}
\bibliography{/Users/Antara/research/latex/apj-jour,/Users/Antara/research/latex/antara_refs,/Users/Antara/research/latex/library_mod,/Users/Antara/research/latex/xray_stacking,/Users/Antara/research/latex/lbgpaper,/Users/Antara/research/UVLG2010/prop_chandra/ms}

\begin{thebibliography}{86}
\expandafter\ifx\csname natexlab\endcsname\relax\def\natexlab#1{#1}\fi

\bibitem[{{Aird} {et~al.}(2010){Aird}, {Nandra}, {Laird}, {Georgakakis},
  {Ashby}, {Barmby}, {Coil}, {Huang}, {Koekemoer}, {Steidel}, \&
  {Willmer}}]{Aird2010}
{Aird}, J., {et~al.} 2010, \mnras, 401, 2531

\bibitem[{{Alexander} {et~al.}(2005){Alexander}, {Bauer}, {Chapman}, {Smail},
  {Blain}, {Brandt}, \& {Ivison}}]{Alexander05}
{Alexander}, D.~M., {et~al.} 2005, \apj, 632, 736

\bibitem[{{Alexander} {et~al.}(2003){Alexander}, {Bauer}, {Brandt},
  {Schneider}, {Hornschemeier}, {Vignali}, {Barger}, {Broos}, {Cowie},
  {Garmire}, {Townsley}, {Bautz}, {Chartas}, \& {Sargent}}]{Alexander03}
---. 2003, \aj, 126, 539

\bibitem[{{Basu-Zych} {et~al.}(in prep)}]{me-chandralba}
{Basu-Zych}, A.~R., {et~al.} in prep

\bibitem[{{Beckwith} {et~al.}(2006){Beckwith}, {Stiavelli}, {Koekemoer},
  {Caldwell}, {Ferguson}, {Hook}, {Lucas}, {Bergeron}, {Corbin}, {Jogee},
  {Panagia}, {Robberto}, {Royle}, {Somerville}, \& {Sosey}}]{Beckwith06}
{Beckwith}, S.~V.~W., {et~al.} 2006, \aj, 132, 1729

\bibitem[{{Belczynski} {et~al.}(2002){Belczynski}, {Kalogera}, \&
  {Bulik}}]{Belczynski2002}
{Belczynski}, K., {Kalogera}, V., \& {Bulik}, T. 2002, \apj, 572, 407

\bibitem[{{Belczynski} {et~al.}(2008){Belczynski}, {Kalogera}, {Rasio}, {Taam},
  {Zezas}, {Bulik}, {Maccarone}, \& {Ivanova}}]{startrack}
{Belczynski}, K., {et~al.} 2008, \apjs, 174, 223

\bibitem[{{Bell} {et~al.}(2005){Bell}, {Papovich}, {Wolf}, {Le Floc'h},
  {Caldwell}, {Barden}, {Egami}, {McIntosh}, {Meisenheimer},
  {P{\'e}rez-Gonz{\'a}lez}, {Rieke}, {Rieke}, {Rigby}, \& {Rix}}]{Bell05}
{Bell}, E.~F., {et~al.} 2005, \apj, 625, 23

\bibitem[{Blain {et~al.}(1999)Blain, Smail, Ivison, \& Kneib}]{Blain99}
Blain, A.~W., Smail, I., Ivison, R.~J., \& Kneib, J.-P. 1999, \mnras, 302, 632

\bibitem[{{Bouwens} {et~al.}(2006){Bouwens}, {Illingworth}, {Blakeslee}, \&
  {Franx}}]{Bouwens06}
{Bouwens}, R.~J., {Illingworth}, G.~D., {Blakeslee}, J.~P., \& {Franx}, M.
  2006, \apj, 653, 53

\bibitem[{{Bouwens} {et~al.}(2007){Bouwens}, {Illingworth}, {Franx}, \&
  {Ford}}]{Bouwens07}
{Bouwens}, R.~J., {Illingworth}, G.~D., {Franx}, M., \& {Ford}, H. 2007, \apj,
  670, 928

\bibitem[{{Bouwens} {et~al.}(2008){Bouwens}, {Illingworth}, {Franx}, \&
  {Ford}}]{Bouwens08}
---. 2008, \apj, 686, 230

\bibitem[{{Bouwens} {et~al.}(2009)}]{Bouwens09}
{Bouwens}, R.~J., {et~al.} 2009, \apj, 705, 936

\bibitem[{{Bouwens} {et~al.}(2010){Bouwens}, {Illingworth}, {Oesch},
  {Stiavelli}, {van Dokkum}, {Trenti}, {Magee}, {Labb{\'e}}, {Franx},
  {Carollo}, \& {Gonzalez}}]{Bouwens10}
---. 2010, \apjl, 709, L133

\bibitem[{{Bouwens} {et~al.}(2011){Bouwens}, {Illingworth}, {Oesch},
  {Labb{\'e}}, {Trenti}, {van Dokkum}, {Franx}, {Stiavelli}, {Carollo},
  {Magee}, \& {Gonzalez}}]{Bouwens11}
---. 2011, \apj, 737, 90

\bibitem[{Bouwens {et~al.}(2011)Bouwens, Illingworth, Oesch, Franx, Labb{\'e},
  Trenti, van Dokkum, Carollo, Gonzalez, \& Magee}]{Bouwens11dust}
Bouwens, R.~J., {et~al.} 2011, arXiv.org, astro-ph.CO, 994

\bibitem[{{Bouwens} {et~al.}(2012){Bouwens}, {Illingworth}, {Oesch}, {Trenti},
  {Labb{\'e}}, {Franx}, {Stiavelli}, {Carollo}, {van Dokkum}, \&
  {Magee}}]{Bouwens2012}
{Bouwens}, R.~J., {et~al.} 2012, \apjl, 752, L5

\bibitem[{{Boylan-Kolchin} {et~al.}(2009){Boylan-Kolchin}, {Springel}, {White},
  {Jenkins}, \& {Lemson}}]{MS-II}
{Boylan-Kolchin}, M., {Springel}, V., {White}, S.~D.~M., {Jenkins}, A., \&
  {Lemson}, G. 2009, \mnras, 398, 1150

\bibitem[{Brandt {et~al.}(2001)Brandt, Hornschemeier, Alexander, Garmire,
  Schneider, Broos, Townsley, Bautz, Feigelson, \& Griffiths}]{Brandt01}
Brandt, W.~N., {et~al.} 2001, The Astronomical Journal, 122, 1

\bibitem[{{Broos} {et~al.}(2011){Broos}, {Getman}, {Povich}, {Townsley},
  {Feigelson}, \& {Garmire}}]{Broos}
{Broos}, P.~S., {et~al.} 2011, \apjs, 194, 4

\bibitem[{{Bunker} {et~al.}(2010){Bunker}, {Wilkins}, {Ellis}, {Stark},
  {Lorenzoni}, {Chiu}, {Lacy}, {Jarvis}, \& {Hickey}}]{Bunker10}
{Bunker}, A.~J., {et~al.} 2010, \mnras, 409, 855

\bibitem[{{Chary} \& {Elbaz}(2001)}]{CharyElbaz}
{Chary}, R., \& {Elbaz}, D. 2001, \apj, 556, 562

\bibitem[{{Colbert} {et~al.}(2004){Colbert}, {Heckman}, {Ptak}, {Strickland},
  \& {Weaver}}]{Colbert04}
{Colbert}, E.~J.~M., {Heckman}, T.~M., {Ptak}, A.~F., {Strickland}, D.~K., \&
  {Weaver}, K.~A. 2004, \apj, 602, 231

\bibitem[{Cowie {et~al.}(2011)Cowie, Barger, \& Hasinger}]{Cowie11}
Cowie, L.~L., Barger, A.~J., \& Hasinger, G. 2011, arXiv.org, 1110, 3326

\bibitem[{{Efron}(1982)}]{DasBoot}
{Efron}, B. 1982, {The Jackknife, the Bootstrap and other resampling plans}
  (CBMS-NSF Regional Conference Series in Applied Mathematics, Philadelphia:
  Society for Industrial and Applied Mathematics (SIAM), 1982)

\bibitem[{{Elbaz} {et~al.}(2011){Elbaz}, {Dickinson}, {Hwang},
  {D{\'{\i}}az-Santos}, {Magdis}, {Magnelli}, {Le Borgne}, {Galliano},
  {Pannella}, {Chanial}, {Armus}, {Charmandaris}, {Daddi}, {Aussel}, {Popesso},
  {Kartaltepe}, {Altieri}, {Valtchanov}, {Coia}, {Dannerbauer}, {Dasyra},
  {Leiton}, {Mazzarella}, {Alexander}, {Buat}, {Burgarella}, {Chary}, {Gilli},
  {Ivison}, {Juneau}, {Le Floc'h}, {Lutz}, {Morrison}, {Mullaney}, {Murphy},
  {Pope}, {Scott}, {Brodwin}, {Calzetti}, {Cesarsky}, {Charlot}, {Dole},
  {Eisenhardt}, {Ferguson}, {F{\"o}rster Schreiber}, {Frayer}, {Giavalisco},
  {Huynh}, {Koekemoer}, {Papovich}, {Reddy}, {Surace}, {Teplitz}, {Yun}, \&
  {Wilson}}]{Elbaz2011}
{Elbaz}, D., {et~al.} 2011, \aap, 533, A119

\bibitem[{Fabbiano(1989)}]{Fabbiano89}
Fabbiano, G. 1989, IN: Annual review of astronomy and astrophysics. Volume 27
  (A90-29983 12-90). Palo Alto, 27, 87

\bibitem[{{Ferguson} {et~al.}(2004){Ferguson}, {Dickinson}, {Giavalisco},
  {Kretchmer}, {Ravindranath}, {Idzi}, {Taylor}, {Conselice}, {Fall},
  {Gardner}, {Livio}, {Madau}, {Moustakas}, {Papovich}, {Somerville},
  {Spinrad}, \& {Stern}}]{Ferguson04}
{Ferguson}, H.~C., {et~al.} 2004, \apjl, 600, L107

\bibitem[{{Fiore} {et~al.}(2012{\natexlab{a}}){Fiore}, {Puccetti}, \&
  {Mathur}}]{Fiore11rev}
{Fiore}, F., {Puccetti}, S., \& {Mathur}, S. 2012{\natexlab{a}}, Advances in
  Astronomy, 2012

\bibitem[{{Fiore} {et~al.}(2012{\natexlab{b}}){Fiore}, {Puccetti}, {Grazian},
  {Menci}, {Shankar}, {Santini}, {Piconcelli}, {Koekemoer}, {Fontana},
  {Boutsia}, {Castellano}, {Lamastra}, {Malacaria}, {Feruglio}, {Mathur},
  {Miller}, \& {Pannella}}]{Fiore11}
{Fiore}, F., {et~al.} 2012{\natexlab{b}}, \aap, 537, A16

\bibitem[{{Fragos} {et~al.}(2012){Fragos}, {Lehmer}, {Tremmel}, {Tzanavaris},
  {Basu-Zych}, {Belczynski}, {Hornschemeier}, {Jenkins}, {Kalogera}, {Ptak}, \&
  {Zezas}}]{F12}
{Fragos}, T., {et~al.} 2012, ArXiv e-prints

\bibitem[{{Franx} {et~al.}(2008){Franx}, {van Dokkum}, {Schreiber}, {Wuyts},
  {Labb{\'e}}, \& {Toft}}]{Franx08}
{Franx}, M., {et~al.} 2008, \apj, 688, 770

\bibitem[{{Gilli} {et~al.}(2011){Gilli}, {Su}, {Norman}, {Vignali}, {Comastri},
  {Tozzi}, {Rosati}, {Stiavelli}, {Brandt}, {Xue}, {Luo}, {Castellano},
  {Fontana}, {Fiore}, {Mainieri}, \& {Ptak}}]{Gilli2011}
{Gilli}, R., {et~al.} 2011, \apjl, 730, L28

\bibitem[{{Grimes} {et~al.}(2006)}]{grimes06}
{Grimes}, J.~P., {et~al.} 2006, \apj, 648, 310

\bibitem[{{Grimes} {et~al.}(2007)}]{grimes07}
---. 2007, \apj, 668, 891

\bibitem[{{Guo} {et~al.}(2011){Guo}, {White}, {Boylan-Kolchin}, {De Lucia},
  {Kauffmann}, {Lemson}, {Li}, {Springel}, \& {Weinmann}}]{Guo2011}
{Guo}, Q., {et~al.} 2011, \mnras, 413, 101

\bibitem[{{Hathi} {et~al.}(2010){Hathi}, {Ryan}, {Cohen}, {Yan}, {Windhorst},
  {McCarthy}, {O'Connell}, {Koekemoer}, {Rutkowski}, {Balick}, {Bond},
  {Calzetti}, {Disney}, {Dopita}, {Frogel}, {Hall}, {Holtzman}, {Kimble},
  {Paresce}, {Saha}, {Silk}, {Trauger}, {Walker}, {Whitmore}, \&
  {Young}}]{Hathi10}
{Hathi}, N.~P., {et~al.} 2010, \apj, 720, 1708

\bibitem[{Hornschemeier {et~al.}(2002)Hornschemeier, Brandt, Alexander, Bauer,
  Garmire, Schneider, Bautz, \& Chartas}]{Hornschemeier02}
Hornschemeier, A.~E., {et~al.} 2002, \apj, 568, 82

\bibitem[{{Iwasawa} {et~al.}(2009){Iwasawa}, {Sanders}, {Evans}, {Mazzarella},
  {Armus}, \& {Surace}}]{iwasawa}
{Iwasawa}, K., {et~al.} 2009, \apjl, 695, L103

\bibitem[{{Kennicutt}(1998)}]{Kennicutt}
{Kennicutt}, R.~C. 1998, \araa, 36, 189

\bibitem[{{Kocevski} {et~al.}(2012){Kocevski}, {Faber}, {Mozena}, {Koekemoer},
  {Nandra}, {Rangel}, {Laird}, {Brusa}, {Wuyts}, {Trump}, {Koo}, {Somerville},
  {Bell}, {Lotz}, {Alexander}, {Bournaud}, {Conselice}, {Dahlen}, {Dekel},
  {Donley}, {Dunlop}, {Finoguenov}, {Georgakakis}, {Giavalisco}, {Guo},
  {Grogin}, {Hathi}, {Juneau}, {Kartaltepe}, {Lucas}, {McGrath}, {McIntosh},
  {Mobasher}, {Robaina}, {Rosario}, {Straughn}, {van der Wel}, \&
  {Villforth}}]{Kocevski}
{Kocevski}, D.~D., {et~al.} 2012, \apj, 744, 148

\bibitem[{{Kroupa}(2001)}]{Kroupa}
{Kroupa}, P. 2001, \mnras, 322, 231

\bibitem[{{Laird} {et~al.}(2005){Laird}, {Nandra}, {Adelberger}, {Steidel}, \&
  {Reddy}}]{Laird05}
{Laird}, E.~S., {Nandra}, K., {Adelberger}, K.~L., {Steidel}, C.~C., \&
  {Reddy}, N.~A. 2005, \mnras, 359, 47

\bibitem[{{Laird} {et~al.}(2006){Laird}, {Nandra}, {Hobbs}, \&
  {Steidel}}]{Laird06}
{Laird}, E.~S., {Nandra}, K., {Hobbs}, A., \& {Steidel}, C.~C. 2006, \mnras,
  373, 217

\bibitem[{{Laird} {et~al.}(2010){Laird}, {Nandra}, {Pope}, \&
  {Scott}}]{Laird2010}
{Laird}, E.~S., {Nandra}, K., {Pope}, A., \& {Scott}, D. 2010, \mnras, 401,
  2763

\bibitem[{{Lee} {et~al.}(2006)}]{Lee06}
{Lee}, K.-S., {et~al.} 2006, \apj, 642, 63

\bibitem[{{Lehmer} {et~al.}(2005{\natexlab{a}}){Lehmer}, {Brandt}, {Alexander},
  {Bauer}, {Schneider}, {Tozzi}, {Bergeron}, {Garmire}, {Giacconi}, {Gilli},
  {Hasinger}, {Hornschemeier}, {Koekemoer}, {Mainieri}, {Miyaji}, {Nonino},
  {Rosati}, {Silverman}, {Szokoly}, \& {Vignali}}]{Lehmer2005ECDFS}
{Lehmer}, B.~D., {et~al.} 2005{\natexlab{a}}, \apjs, 161, 21

\bibitem[{{Lehmer} {et~al.}(2005{\natexlab{b}}){Lehmer}, {Brandt}, {Alexander},
  {Bauer}, {Conselice}, {Dickinson}, {Giavalisco}, {Grogin}, {Koekemoer},
  {Lee}, {Moustakas}, \& {Schneider}}]{L05}
---. 2005{\natexlab{b}}, \aj, 129, 1 (L05)

\bibitem[{{Lehmer} {et~al.}(2007){Lehmer}, {Brandt}, {Alexander}, {Bell},
  {McIntosh}, {Bauer}, {Hasinger}, {Mainieri}, {Miyaji}, {Schneider}, \&
  {Steffen}}]{Lehmer07}
---. 2007, \apj, 657, 681

\bibitem[{{Lehmer} {et~al.}(2008)}]{L08}
---. 2008, \apj, 681, 1163 (L08)

\bibitem[{{Lehmer} {et~al.}(2010)}]{Lehmer2010}
---. 2010, \apj, 724, 559

\bibitem[{{Lehmer} {et~al.}(2012){Lehmer}, {Xue}, {Brandt}, {Alexander},
  {Bauer}, {Brusa}, {Comastri}, {Gilli}, {Hornschemeier}, {Luo}, {Paolillo},
  {Ptak}, {Shemmer}, {Schneider}, {Tozzi}, \& {Vignali}}]{Lehmer12}
---. 2012, \apj, 752, 46

\bibitem[{{Madau} {et~al.}(1996){Madau}, {Ferguson}, {Dickinson}, {Giavalisco},
  {Steidel}, \& {Fruchter}}]{Madau96}
{Madau}, P., {et~al.} 1996, \mnras, 283, 1388

\bibitem[{{Meurer} {et~al.}(1999){Meurer}, {Heckman}, \& {Calzetti}}]{Meurer99}
{Meurer}, G.~R., {Heckman}, T.~M., \& {Calzetti}, D. 1999, \apj, 521, 64

\bibitem[{{Miller} {et~al.}(2012)}]{Millerinprep}
{Miller}, N., {et~al.} 2012, in prep.

\bibitem[{{Miller} {et~al.}(2008){Miller}, {Fomalont}, {Kellermann},
  {Mainieri}, {Norman}, {Padovani}, {Rosati}, \& {Tozzi}}]{Miller08}
{Miller}, N.~A., {et~al.} 2008, \apjs, 179, 114

\bibitem[{{Mineo} {et~al.}(2012){Mineo}, {Gilfanov}, \& {Sunyaev}}]{Mineo12}
{Mineo}, S., {Gilfanov}, M., \& {Sunyaev}, R. 2012, \mnras, 419, 2095

\bibitem[{{Mosleh} {et~al.}(2011){Mosleh}, {Williams}, {Franx}, \&
  {Kriek}}]{Mosleh11}
{Mosleh}, M., {Williams}, R.~J., {Franx}, M., \& {Kriek}, M. 2011, \apj, 727, 5

\bibitem[{{Mullaney} {et~al.}(2012){Mullaney}, {Pannella}, {Daddi},
  {Alexander}, {Elbaz}, {Hickox}, {Bournaud}, {Altieri}, {Aussel}, {Coia},
  {Dannerbauer}, {Dasyra}, {Dickinson}, {Hwang}, {Kartaltepe}, {Leiton},
  {Magdis}, {Magnelli}, {Popesso}, {Valtchanov}, {Bauer}, {Brandt}, {Del Moro},
  {Hanish}, {Ivison}, {Juneau}, {Luo}, {Lutz}, {Sargent}, {Scott}, \&
  {Xue}}]{JM12}
{Mullaney}, J.~R., {et~al.} 2012, \mnras, 419, 95

\bibitem[{{Oesch} {et~al.}(2010{\natexlab{a}}){Oesch}, {Bouwens}, {Carollo},
  {Illingworth}, {Magee}, {Trenti}, {Stiavelli}, {Franx}, {Labb{\'e}}, \& {van
  Dokkum}}]{Oesch2010}
{Oesch}, P.~A., {et~al.} 2010{\natexlab{a}}, \apjl, 725, L150

\bibitem[{{Oesch} {et~al.}(2010{\natexlab{b}}){Oesch}, {Bouwens},
  {Illingworth}, {Carollo}, {Franx}, {Labb{\'e}}, {Magee}, {Stiavelli},
  {Trenti}, \& {van Dokkum}}]{Oesch10}
---. 2010{\natexlab{b}}, \apjl, 709, L16

\bibitem[{{Oesch} {et~al.}(2012){Oesch}, {Bouwens}, {Illingworth}, {Labb{\'e}},
  {Trenti}, {Gonzalez}, {Carollo}, {Franx}, {van Dokkum}, \&
  {Magee}}]{Oesch2012b}
---. 2012, \apj, 745, 110

\bibitem[{{Persic} \& {Rephaeli}(2007)}]{PR2007}
{Persic}, M., \& {Rephaeli}, Y. 2007, \aap, 463, 481

\bibitem[{Ptak {et~al.}(2001)Ptak, Griffiths, White, \& Ghosh}]{Ptak01}
Ptak, A., Griffiths, R., White, N., \& Ghosh, P. 2001, \apj, 559, L91

\bibitem[{{Ptak} {et~al.}(2007){Ptak}, {Mobasher}, {Hornschemeier}, {Bauer}, \&
  {Norman}}]{Ptak07}
{Ptak}, A., {Mobasher}, B., {Hornschemeier}, A., {Bauer}, F., \& {Norman}, C.
  2007, ArXiv e-prints

\bibitem[{{Ptak} {et~al.}(1999){Ptak}, {Serlemitsos}, {Yaqoob}, \&
  {Mushotzky}}]{Ptak99}
{Ptak}, A., {Serlemitsos}, P., {Yaqoob}, T., \& {Mushotzky}, R. 1999, \apjs,
  120, 179

\bibitem[{{Ranalli} {et~al.}(2003){Ranalli}, {Comastri}, \&
  {Setti}}]{Ranalli03}
{Ranalli}, P., {Comastri}, A., \& {Setti}, G. 2003, \aap, 399, 39

\bibitem[{{Reddy} \& {Steidel}(2004)}]{Reddy04}
{Reddy}, N.~A., \& {Steidel}, C.~C. 2004, \apjl, 603, L13

\bibitem[{{Reddy} \& {Steidel}(2009)}]{Reddy09}
---. 2009, \apj, 692, 778

\bibitem[{{Reddy} {et~al.}(2006)}]{Reddy}
{Reddy}, N.~A., {et~al.} 2006, \apj, 653, 1004

\bibitem[{{Silverman} {et~al.}(2008){Silverman}, {Green}, {Barkhouse}, {Kim},
  {Kim}, {Wilkes}, {Cameron}, {Hasinger}, {Jannuzi}, {Smith}, {Smith}, \&
  {Tananbaum}}]{Silverman08}
{Silverman}, J.~D., {et~al.} 2008, \apj, 679, 118

\bibitem[{{Stark} {et~al.}(1992){Stark}, {Gammie}, {Wilson}, {Bally}, {Linke},
  {Heiles}, \& {Hurwitz}}]{Stark92}
{Stark}, A.~A., {et~al.} 1992, \apjs, 79, 77

\bibitem[{{Steidel} \& {Hamilton}(1992)}]{Steidel92}
{Steidel}, C.~C., \& {Hamilton}, D. 1992, \aj, 104, 941

\bibitem[{{Steidel} \& {Hamilton}(1993)}]{Steidel93}
---. 1993, \aj, 105, 2017

\bibitem[{{Steidel} {et~al.}(1995){Steidel}, {Pettini}, \&
  {Hamilton}}]{Steidel95}
{Steidel}, C.~C., {Pettini}, M., \& {Hamilton}, D. 1995, \aj, 110, 2519

\bibitem[{{Steidel} {et~al.}(2000)}]{Steidel00}
{Steidel}, C.~C., {et~al.} 2000, \apj, 532, 170

\bibitem[{{Symeonidis} {et~al.}(2011){Symeonidis}, {Georgakakis}, {Seymour},
  {Auld}, {Bock}, {Brisbin}, {Buat}, {Burgarella}, {Chanial}, {Clements},
  {Cooray}, {Eales}, {Farrah}, {Franceschini}, {Glenn}, {Griffin},
  {Hatziminaoglou}, {Ibar}, {Ivison}, {Mortier}, {Oliver}, {Page},
  {Papageorgiou}, {Pearson}, {P{\'e}rez-Fournon}, {Pohlen}, {Rawlings},
  {Raymond}, {Rodighiero}, {Roseboom}, {Rowan-Robinson}, {Scott}, {Smith},
  {Tugwell}, {Vaccari}, {Vieira}, {Vigroux}, {Wang}, \&
  {Wright}}]{Symeonidis11}
{Symeonidis}, M., {et~al.} 2011, \mnras, 417, 2239

\bibitem[{{Thompson} {et~al.}(2005){Thompson}, {Illingworth}, {Bouwens},
  {Dickinson}, {Eisenstein}, {Fan}, {Franx}, {Riess}, {Rieke}, {Schneider},
  {Stobie}, {Toft}, \& {van Dokkum}}]{Thompson}
{Thompson}, R.~I., {et~al.} 2005, \aj, 130, 1

\bibitem[{Treister {et~al.}(2010)Treister, Natarajan, Sanders, Urry,
  Schawinski, \& Kartaltepe}]{Treister11}
Treister, E., {et~al.} 2010, arXiv, astro-ph.CO

\bibitem[{{Trujillo} {et~al.}(2006){Trujillo}, {F{\"o}rster Schreiber},
  {Rudnick}, {Barden}, {Franx}, {Rix}, {Caldwell}, {McIntosh}, {Toft},
  {H{\"a}ussler}, {Zirm}, {van Dokkum}, {Labb{\'e}}, {Moorwood},
  {R{\"o}ttgering}, {van der Wel}, {van der Werf}, \& {van
  Starkenburg}}]{Trujillo06}
{Trujillo}, I., {et~al.} 2006, \apj, 650, 18

\bibitem[{{Ueda} {et~al.}(2003){Ueda}, {Akiyama}, {Ohta}, \& {Miyaji}}]{Ueda03}
{Ueda}, Y., {Akiyama}, M., {Ohta}, K., \& {Miyaji}, T. 2003, \apj, 598, 886

\bibitem[{{Watson} {et~al.}(2009){Watson}, {Kochanek}, {Forman}, {Hickox},
  {Jones}, {Brown}, {Brand}, {Dey}, {Jannuzi}, {Kenter}, {Murray}, {Vikhlinin},
  {Eisenstein}, {Fazio}, {Green}, {McNamara}, {Rieke}, \& {Shields}}]{Watson09}
{Watson}, C.~R., {et~al.} 2009, \apj, 696, 2206

\bibitem[{Willott(2011)}]{Willott11}
Willott, C.~J. 2011, arXiv.org, astro-ph.CO

\bibitem[{{Xue} {et~al.}(2011){Xue}, {Luo}, {Brandt}, {Bauer}, {Lehmer},
  {Broos}, {Schneider}, {Alexander}, {Brusa}, {Comastri}, {Fabian}, {Gilli},
  {Hasinger}, {Hornschemeier}, {Koekemoer}, {Liu}, {Mainieri}, {Paolillo},
  {Rafferty}, {Rosati}, {Shemmer}, {Silverman}, {Smail}, {Tozzi}, \&
  {Vignali}}]{X11}
{Xue}, Y.~Q., {et~al.} 2011, \apjs, 195, 10 (X11)

\bibitem[{{Xue} {et~al.}(2012)}]{Xue12}
---. 2012, submitted.

\bibitem[{{Yun} {et~al.}(2001){Yun}, {Reddy}, \& {Condon}}]{Yun}
{Yun}, M.~S., {Reddy}, N.~A., \& {Condon}, J.~J. 2001, \apj, 554, 803

\end{thebibliography}

\end{document}